\documentclass[manuscript]{aastex}
\shorttitle{Herschel: 3C~454.3}
\shortauthors{Wehrle et al.}

\begin{document}

\title{Multiwavelength Variations of 3C~454.3 during the November 2010 to January 2011 Outburst}

\author{Ann E. Wehrle\altaffilmark{1}, 
Alan P. Marscher\altaffilmark{2}, Svetlana G. Jorstad\altaffilmark{2,3}, Mark A. Gurwell\altaffilmark{4},
Manasvita Joshi\altaffilmark{2}, Nicholas R. MacDonald\altaffilmark{2},
Karen E. Williamson\altaffilmark{2}, Iv\'an Agudo\altaffilmark{5,2},
and Dirk Grupe\altaffilmark{6}
}
\altaffiltext{1}{Space Science Institute, 4750 Walnut Street, Suite 205, Boulder, CO 80301}
\email{awehrle@spacescience.org}
\altaffiltext{2}{Boston University, Department of Astronomy, 725 Commonwealth Avenue, Boston MA 02215}
\altaffiltext{3}{Astronomical Institute, St. Petersburg State University, Universitetskij Pr. 28, Petrodvorets, 
198504 St. Petersburg, Russia}
\altaffiltext{4}{Harvard-Smithsonian Center for Astrophysics, MS 42, 60 Garden St., Cambridge, MA 02138}
\altaffiltext{5}{Instituto de Astrof\'{\i}sica de Andaluc\'{\i}a, CSIC, Apartado 3004, 18080,
Granada, Spain}
\altaffiltext{6}{Department of Astronomy and Astrophysics, Pennsylvania State University, 525 Davey Lab, University Park, PA 16802}

\begin{abstract}
We present multiwavelength data of the blazar 3C~454.3 obtained during an extremely bright outburst from November 2010 through January 2011. These include flux density measurements with the {\it Herschel Space Observatory} at five submillimeter-wave and far-infrared bands, the {\it Fermi} Large Area Telescope at $\gamma$-ray energies, {\it Swift} at X-ray, ultraviolet (UV), and optical frequencies, and the Submillimeter Array at 1.3 mm. From this dataset, we form a series of 52 spectral energy distributions (SEDs) spanning nearly two months that are unprecedented in time coverage and breadth of frequency. Discrete correlation anlaysis of the millimeter, far-infrared, and $\gamma$-ray light curves show that the variations were essentially simultaneous, indicative of co-spatiality of the emission, at these wavebands. In contrast, differences in short-term fluctuations at various wavelengths imply the presence of inhomegeneities in physical conditions across the source. We locate the site of the outburst in the parsec-scale ``core'', whose flux density as measured 
on 7 mm Very Long Baseline Array images increased by 70\% during the first five weeks of the outburst. Based on these considerations and guided by the SEDs, we propose a model in which turbulent plasma crosses a conical standing shock in the parsec-scale region of the jet. Here, the high-energy 
emission in the model is produced by inverse Compton scattering of seed photons supplied by either
nonthermal radiation from a Mach disk, thermal emission from hot dust, or (for X-rays) synchrotron
radiation from plasma that crosses the standing shock. For the two dates on which we fitted the model SED to the data, the model
corresponds very well to the observations at all bands except at X-ray
energies, where the spectrum is flatter than observed.
\end{abstract}


\section{Introduction}

The $\gamma$-ray sky at high Galactic latitudes is dominated by highly variable active galactic nuclei termed ``blazars.'' Extreme $\gamma$-ray apparent luminosities that can exceed $10^{48}$ erg s$^{-1}$ during outbursts in blazars are most
easily understood as the result of Doppler boosting of the nonthermal emission from a relativistic
plasma jet pointing within several degrees of our line of sight \citep[e.g.][]{dermer95}. The Doppler effect also shortens the observed
timescale of variability and accentuates the amplitude of the flux changes. The emission at radio to optical --- and in some cases up to X-ray --- frequencies is synchrotron radiation from ultra-relativistic electrons
gyrating in magnetic fields inside the jet.

The electrons that give rise to the synchrotron emission also create X-rays and $\gamma$-rays
through inverse Compton scattering. The synchrotron radiation, as well as other types of emission from 
various regions in the galactic nucleus, can provide seed photons for scattering.
Neither the main source(s) of these seed photons nor
the processes that energize the radiating electrons have yet been identified with certainty. A promising
method for doing so is to observe and model both the spectral energy distribution (SED) at various stages of
an outburst and the time delays between variations at millimeter, infrared (IR), 
optical, ultraviolet (UV), X-ray, and $\gamma$-ray bands during outbursts.
The goal is to use the results of such an
analysis to identify the location of the scattering regions (i.e., in the vicinity of the accretion disk
$\lesssim 10^{17}$ cm from the central supermassive black hole, within the inner parsec, or farther out)
and, by doing so, infer the source of the seed photons [nonthermal emission from the jet or thermal
emission from the accretion disk, broad emission-line region (BLR), or parsec-scale hot dust].
This information can
then be used to define, or place strong constraints on, the physical conditions in the inner $\sim 10$ pc
of the jet (e.g., magnetic field, electron energy density, energy gains and losses of the relativistic 
electrons, and changes in bulk Lorentz factor). 

The flat-radio-spectrum quasar 3C~454.3 (redshift of 0.859) is particularly well-suited for such a
study. Analysis of time
sequences of images with angular resolution of $\sim 0.1$ milliarcseconds (mas) with the Very Long Baseline
Array (VLBA) at a frequency of 43 GHz shows that this blazar
contains a narrow (opening half-angle $\lesssim 1^\circ$) relativistic jet pointed within $2^\circ$ of our line of sight \citep{J05}. An extremely bright multiwavelength outburst in 2005 provided an example of the 
high-amplitude variability of the jet \citep{Vil06,J10}.

On 2010 October 31, observers discovered that 3C~454.3 was undergoing a pronounced flare at near-IR
wavelengths \citep{carrasco2010}. The flare extended to millimeter, IR, optical, UV,
X-ray, and $\gamma$-ray bands  \citep[see][and references therein]{vercellone2011}.
In 2010 November, 3C~454.3 attained the highest flux ever observed at $\gamma$-ray energies near 1 GeV
\citep{vercellone2010,vercellone2011,striani2010,sanchez2010,abdo2011}, peaking on November 19-20.    
Based on the extraordinary nature of the outburst, we obtained a series of {\it Herschel} and {\it Swift} 
target of opportunity observations of the quasar\footnote{{\it Herschel} is a European Space Agency space observatory with science instruments provided by European-led principal investigator consortia and with important participation from NASA.}.
This was the first time that {\it far}-IR, X-ray, and $\gamma$-ray space telescopes were all available to observe
simultaneously during a truly extraordinary blazar outburst. We have previously
observed flares of blazars at mid-IR and near-IR bands with the {\it Spitzer Space Telescope}, first without and then with concurrent {\it AGILE} $\gamma$-ray observations \citep[][ Wehrle et al., in preparation]{ogle2011}. In those observations, we unexpectedly discovered that there were two peaks in the synchrotron SED. The synchrotron SED of BL Lac has also been found to contain a double hump profile at higher frequencies \citep{raiteri2010}. Such structure in the SED could imply the existence of two distinct sites of IR emission in the jet. One possibility is that both a relatively
quiescent component --- e.g., the quasi-stationary ``core'' seen in millimeter-wave VLBA images --- and
a propagating disturbance --- appearing as a superluminal knot in sequences of such images --- are present during
outbursts. On the other hand, during a singularly bright flare a new knot should dominate the flux, producing
a single peak in the SED.  Two peaks in the synchrotron SED can also be explained by two separate electron-positron populations. 
Hadronic proton-proton and proton-photon collisions could make enough pions to produce, via decays and electromagnetic cascading, a secondary electron-positron population.
 Alternatively, two electron acceleration mechanisms, e.g., magnetic reconnections and shocks, could operate, or  the physical conditions could vary across the source.

\citet{J10} have linked strong $\gamma$-ray flares with
the appearance of superluminal knots at millimeter wavelengths in 3C~454.3. If this is the case for all
such events, we expect to see increased activity in the jet in VLBA images at 7 mm that
we obtained during and after the outburst.

Here we present extensive multiwavelength data during the outburst in 3C~454.3 and discuss
its implications. Our data extend the results presented by \citet{abdo2011} and \citet{vercellone2011},
especially through the addition of {\it Herschel} and VLBA observations. We also offer a different
interpretation of the location of the event and the source of seed photons that were scattered up to
$\gamma$-ray energies. When translating angular to linear sizes, we adopt the current standard
flat-spacetime cosmology, with Hubble constant $H_0$=71 km s$^{-1}$ Mpc$^{-1}$,
$\Omega_M=0.27$, and $\Omega_\Lambda=0.73$. At a redshift $z=0.859$, 3C~454.3 has a
luminosity distance $d_\ell$ = 5.489 Gpc, and 1 mas corresponds to a projected distance
of 7.7 pc in the rest frame of the host galaxy of the quasar.

\section{Observations}

Our {\it Herschel} target of opportunity observations began on 2010 November 19 (MJD 55519). Near-daily {\it Swift}  pointings were arranged by us as well as by other groups starting on 2010 November 2, with a few gaps caused by moon avoidance, $\gamma$-ray burst observations, and other incidents. Observations at other wavebands were already underway or were initiated shortly thereafter. The Large Area Telescope (LAT) instrument on {\it Fermi} operated in standard 
survey mode, scanning the entire sky every three hours. Twice-weekly observations of 3C~454.3 were carried out at the Submillimeter Array (SMA) on Mauna Kea, HI. VLBA observations took place on 2010 November 1, 6, and 13 and December 4
as part of the ongoing Boston University (BU) $\gamma$-ray blazar monitoring program.
We obtained limited mid-IR flux measurements of 3C~454.3 with the NASA
Infrared Telescope Facility (IRTF) on Mauna Kea on 2010 November 3, during the early stage of the
outburst.
 
\subsection{Herschel Observations}
Our {\it Herschel} observing plan was designed to obtain
both time delays and spectral energy distributions with intensive observing on every possible day when
the PACS-Photometer and SPIRE-Photometer modes were operating during
the 51-day visibility window.  PACS and SPIRE observations are normally carried out in separate instrument campaigns 10-14 days long; on changeover days, both instruments can observe within a 24 hour period.  

Each PACS-Photometer observation was carried out in ``scan map" mode with 3-arcminute legs and a cross-scan step of 4 arcseconds with a total on-source time of 72 seconds. Each PACS-Photometer observation included ``70 $\mu$m" (60-85 $\mu$m bandpass) and ``160 $\mu$m" (130-210 $\mu$m bandpass) bands. The SPIRE-Photometer observations were done in ``small map" mode with a total on-source time of 37 seconds. We obtained a total of 42 PACS-Photometer images at each filter band (70, 160 $\mu$m) taken on 15 days between 2010 Nov 19 and 2011 January 10. Similarly, we obtained a total of 13 SPIRE-Photometer images at each filter band (250, 350, 500 $\mu$m) taken on 13 days between 2010 November 23 and 2011 January 9. On 2010 Dec 7 (MJD 55537) and 2011 Jan 8 (MJD 55569), we obtained both SPIRE and PACS data. We derived flux densities from the Herschel pipeline images (version 6.1) from aperture photometry carried out in the Herschel Interactive Processing Environment. Annular sky photometry using HIPE task ``annularSkyAperturePhotometry" was used for PACS images, and aperture corrections applied as tabulated in 
the ``PACS Photometer- Point Source Flux Calibration'' document (version 1.0, 12 April 2011, Herschel Document PICC-ME-TN-037), Table 14. The PACS systematic error was 2.64\% at 70 $\mu$m and 4.15\% at
160 $\mu$m (PACS Photometer -Point Source Calibration, p. 23).

For the SPIRE measurements, we used Gaussian-fitting photometry on {\it Herschel} pipeline images (version 6.1), via HIPE task ``sourceFitting'', and then applied aperture and pixelization corrections as described in ``The SPIRE Photometry Cookbook'' (version 3 May 2011). Although the SPIRE fields are crowded with very faint sources --- probably foreground galaxies --- none were bright enough to significantly contaminate the aperture photometry measurements of the very bright quasar. The SPIRE systematic error was 5\% (SPIRE Photometry Cookbook, version 3 May 2011, and references therein).

\begin{figure}
\plotone{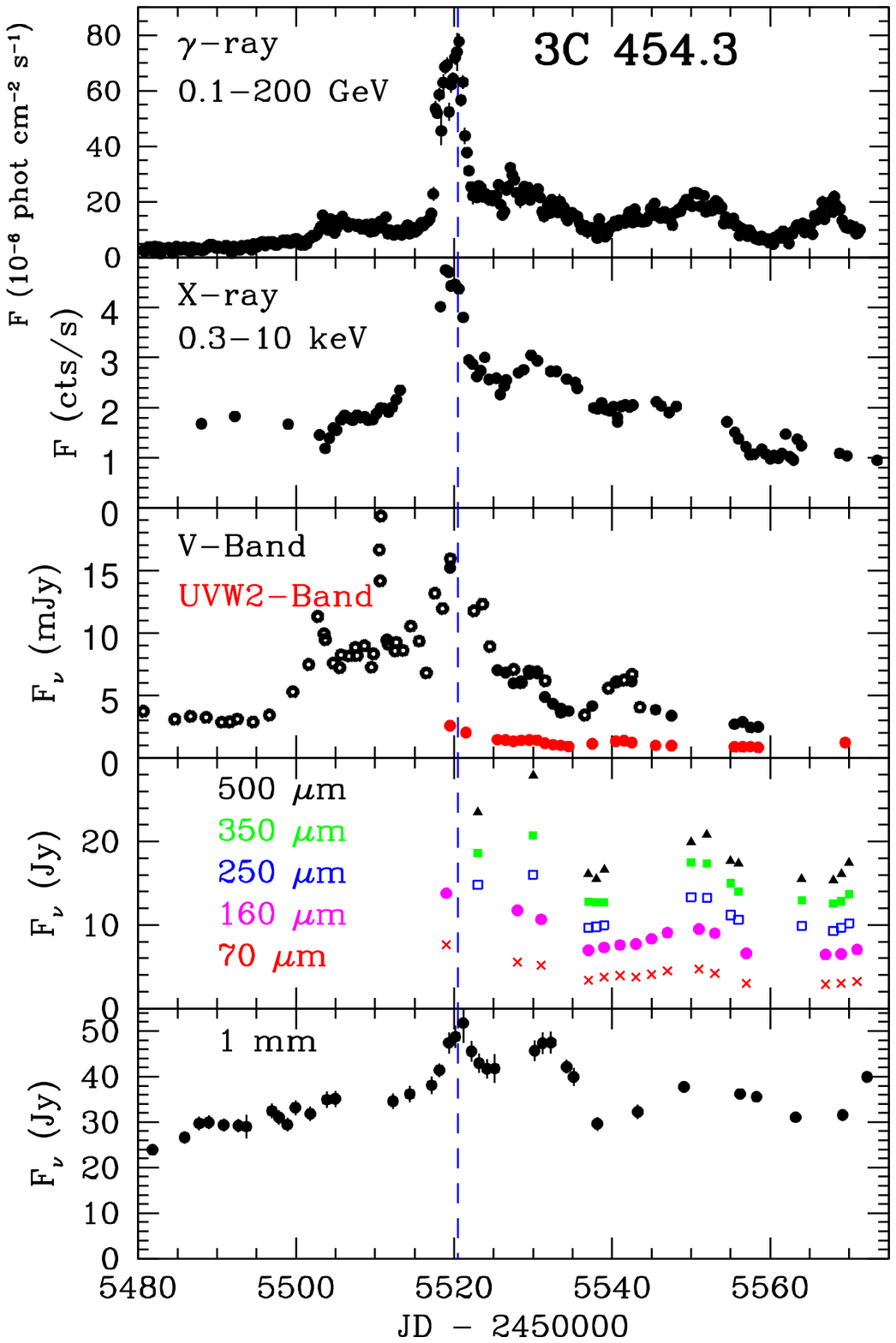}
\caption{Light curves during November 2010-January 2011. From top: Fermi LAT $\gamma-$ray; {\it Swift} X-ray; V-band (filled circles are Swift
V-band, open circles from SMARTS and Perkins
Telescope); {\it Swift} UVW2 band (filled red circles); {\it Herschel} 500 (black triangles), 350 (filled green squares), 250 (open purple squares), 160 (filled magenta circles) and 70 (red crosses) $\mu$m bands; and SMA 1.3 mm band. The vertical dashed line is set at MJD 55520, the peak of the $\gamma-$ ray flare.
\label{fig1}}
\end{figure}

In some cases, we had several observations on the same day taken a few minutes apart in different scan directions or a few hours apart. We reduced each observation separately. The intra-day measurement differences were less than 2\%, and in most cases less than 1\%, well within the systematic errors; 
consequently, we averaged the data for each day. No color correction was applied (equivalent to assuming that $F_\nu \propto \nu^{-\alpha}$, with $\alpha=1$, or $\nu F_\nu=constant$ in SPIRE and PACS terminologies),
since we did not know, a priori, the value of the spectral index over the {\it Herschel} bandpasses.
We present the results in {\bf Table 1}  and the light curves in
Figure \ref{fig1}.

\subsection{Submillimeter Array and IRTF Observations}

The quasar 3C~454.3 is commonly used as an amplitude and phase calibrator at the SMA. Flux history measurements
at wavelengths of 1.3~mm, 850 $\mu$m, and 450 $\mu$m of 3C~454.3 and several hundred other blazars in the Submillimeter Calibrator List, maintained by M. Gurwell, are provided online at http://sma1.sma.hawaii.edu. During the 2010 November to 2011 January period, we also observed the quasar as a science target.  Data were reduced in the usual manner, as described in \citet{gurwell07}. Data obtained at 225 GHz (1.3 mm) during the {\it Herschel} observing period are listed in {\bf Table 1}.  

The IRTF observations were performed with the MIRSI camera \citep{kassis08} at central wavelengths
of 4.9, 10.6, and 20.7 $\mu$m. We reduced the IRTF data with IDL using a script supplied by
the IRTF staff, with calibration based on standard stars from the IRTF catalog.

\subsection{Perkins Telescope and SMARTS Observations}

We obtained optical photometric data at the 1.8 m Perkins Telescope of
 Lowell Observatory (Flagstaff, AZ) in BVR bands. We have
employed differential photometry with comparison
stars provided by \citet{RAI98}.  The V-band optical data are supplemented by publicly available measurements by the SMARTS consortium (Charles Bailyn, PI), posted at their website http://www.astro.yale.edu/smarts/glast/.  The V-band data are shown in Figure 1. Additional data and analysis will be given in Jorstad et al. (2012).

\subsection{Swift Optical, Ultraviolet, and X-ray Observations}
 
We analyzed all available data obtained by the Swift UVOT during the outburst that produced measurements of the flux density of 3C 454.3 in six filters, V (5402~\AA), B (4329~\AA), U (3501~\AA), UW1 (2634~\AA), UM2 (2231~\AA), and UW2 (2030~\AA). We employed the UVOT software task {\it uvotsource} to extract counts within a circular region of 5 arcsec radius for the source and 10 arcsec radius for the background, the latter centered 30 arcsec south-west of the source.
We used the coefficients given in Table 1 of \citet{raiteri2011} for correcting the derived magnitudes and flux densities for Galactic extinction. The de-reddened data are listed in {\bf Tables 2-4}.

We used version 3.8 of the {\it Swift} Software (HEAsoft 6.11),
with the version of CALDB released on 2011 June 9, to reduce the XRT data collected for 3C~454.3. During the outburst, from MJD 55457 to MJD 
55578, the XRT obtained 67 measurements in Windowed Timing mode (WT) and 28 measurements in Photon Counting mode (PC). Observations in WT mode were performed during the brightest flux levels to avoid a pile-up effect on images, since from 2010 November 9 (MJD 55509) to December 24 (MJD 55554) the source count rate exceeded 1~count~s$^{-1}$ at 0.3-10~keV and reached a maximum of 4.028$\pm$0.008~count~s$^{-1}$ on November 19. We have downloaded Level 2 event files from the {\it Swift} Database Archive 
and reduced them in the manner described in the {\it SWIFT XRT Data Reduction Guide}, available on the {\it Swift Data Analysis} web page. The task {\it xselect} was used for extracting spectra of the source and background. In PC mode we used a circular aperture of 30 pixels for the source and an annulus with inner/outer radii of 110/160 pixels for the background. In WT mode, rectangular apertures equal in size to the circular and annular ones were applied along the brightest part of the window for the source and at the end of the window for the background. The tasks {\it xrtmkarf} and {\it grppha} were employed to produce the effective area file and to rebin the energy
channels, respectively. Rebinned spectra were modelled within 0.3-10~keV in XSPEC v.12.7 with a single power-law and Galactic absorption corresponding to a hydrogen column density N$_H$ = 6.50$\times$10$^{20}$~cm$^{-2}$ \citep{kalberla2005}. We applied a Monte-Carlo method to estimate the goodness of fit of the model and derived 90\% ($\sim$3-$\sigma$) confidence ranges of the fitted parameters (photon index and amplitude).  The X-ray data are listed in {\bf Table 5}.

\subsection{Fermi LAT Data Analysis}

We downloaded Pass 7 {\it Fermi} LAT data for the field surrounding
3C~454.3 from the data server of the Fermi Science Support Center (FSSC)
at URL http://fermi.gsfc.nasa.gov/ssc. Reduction of the data utilized
version v9r23p1 of the Fermi Science Tools software, associated background
models, and a model of sources in the field (within $\sim 20^\circ$ of the
position of 3C~454.3) generated from the 2FGL catalog by the script
``make2FGLxml.py" by T. Johnson, available at the same
website. We used the standard unbinned maximum likelihood analysis, integrating over short,
6-hour time bins and a photon energy
range of 0.1-200 GeV. The $\gamma$-ray emission was detected at a test statistic value $> 5$
\citep[equivalent to about 2-$\sigma$; see][]{Mat96} at all but one of the time intervals.
The spectral model for 3C~454.3 was fixed to that given in the 2FGL
catalog: a log-parabolic shape of the photon flux spectrum as a function
of energy $E_\gamma$:
\begin{equation}
F(E_\gamma) = F_0 (E_\gamma/E_b)^{-(\alpha+\beta\ln(E_\gamma/E_b))},
\end{equation}
with $\alpha=2.23$, $\beta=0.118$, and $E_b =298$ MeV.
We used a similar analysis for 24-hour binned data, except that we allowed the maximum likelihood algorithm
to find the optimal values of $\alpha$, $\beta$, and $E_b$ rather than fixing them. The 24-hour-binned data are listed in {\bf Table 6}.
 Because the errors in the log-parabolic model are correlated, their behavior is not simple, and we refer the reader to the technical discussion in the appendix of \citet{tramacere2007}. We have adopted the standard deviations of the $\alpha$, $\beta$ and break energies as uncertainties. 
Specifically,  the errors are estimated at 6\% in $\alpha$ and 15\% for $\beta$ when the photon flux exceeds $10^{-5}$~ $ph/s/cm^{2}$ and 35\% below that level, and 30\% in the break energy.
(Running the maximum likelihood code on 24-hour binned data three times with different random number seeds resulted in changes of 1-2\% in the values of alpha and beta.) There were too few photons when the data were binned in 24-hour bins to determine the break energy to better than about 30\%. Our results are in general agreement with those of \citet{abdo2011} who divided the flare episode into four unequal periods of several-to-many days each to obtain improved statistics using more photons. An additional contributor to the smaller error bars obtained by Abdo et al. follows from their use of a constant break energy while we allowed the break energy to vary.

\subsection{VLBA Imaging}

We observed 3C~454.3 with the VLBA at 43 GHz in a manner similar to that described by \citet{J10}, where details of the
data calibration and imaging can be found. Figure \ref{fig3} presents four images from epochs within one month of
the date of the maximum $\gamma$-ray flux during the outburst. The image features the bright, compact ``core'' A0 at
the eastern end of the jet, as well as superluminal knot K09 that passed through the core during the late 2009 outburst.
Observations with the VLBA in 2011 reveal that a new superluminal knot, K10 with an apparent speed of $\sim 9c$, was
passing through the core during 2010 November \citep[see][]{jorstad2012}. The outburst in
3C~454.3 therefore provides us with an unprecedented opportunity to follow the evolution of a well-sampled 
SED as a superluminal knot traverses different regions of the jet.

\section{Results}
\subsection{Light Curves and Spectral Indices}

\subsubsection {Herschel}

The {\it Herschel} light curves are shown in  Figure \ref{fig1}. The observations began on MJD 55519, yielding
maximum flux densities of 13.8 and 7.61 Jy at 160 and 70 $\mu$m, respectively. The highest flux density was observed at 500 $\mu$m, on MJD 55530. The brightness decreased after this, interrupted by a brief rise on MJD 55550-55552 followed
by further decline, with a slight upturn at the end of the observing window at MJD 55570-55571. Band-to-band spectral indices in the 1.3 mm - 70 $\mu$m range varied significantly, as measured on individual days. The overall variation was a factor of 1.8 at 500 $\mu$m, 1.7 at 350 $\mu$m, 1.7 at 250 $\mu$m, 2.1 at 160 $\mu$m, and 2.6 at 70 $\mu$m. The spectral indices, standard deviation and number of measurements
(in parentheses) are:
1.3mm-500 $\mu$m: $0.64 \pm 0.06 (5)$, 500-350 $\mu$m: $0.59 \pm 0.12 (13)$, 350-250 $\mu$m: $0.81 \pm 0.06,  (13)$, 350-160 $\mu$m: $0.77 \pm 0.10 (3)$, 160-70 $\mu$m: $0.88 \pm 0.07 (14)$. Individual errors on the spectral indices are 2-4\%.

\subsubsection  {SMA and IRTF}

During the weeks leading up to the brightest days of the flare, the 1.3 mm flux density increased from a minimum of 3 Jy in April 2009 to a plateau of about 20-30 Jy, then began rising abruptly on  MJD 55515. It peaked on MJD 55520 at $51.8\pm 4.3$ Jy, then declined to $41.8  \pm 3.1$ Jy over the next few days. A brief flare to $47.5 \pm 2.4$ Jy occurred on MJD 55529-55531, coincident with a high flux at 500 $\mu$m observed by {\it Herschel}.
The mid-IR flux densities that we measured with the IRTF on 3 November 2010 were
$0.293\pm 0.021$ Jy, $0.699\pm 0.037$ Jy, and $1.293\pm 0.195$ Jy  at 4.9, 10.6, and 20.7 $\mu$m, respectively,
so that the corresponding spectral indices were $\alpha_{4.9-10.6} = 1.13\pm 0.11$ and $\alpha_{20.7-10.6} = 0.92\pm 0.13$.

\subsubsection {Swift UVOT and XRT}
The {\it Swift} UVOT light curves in V, B, U, UVW1, UVM2, and UVW2 bands are presented in Figure \ref{fig4};  with additional V-band data from the Perkins Observatory and the SMARTS program. Inspection of this plot reveals that the main flare on MJD 55519-20 was preceded by a brightness plateau of abut two weeks, followed by two significant flares. The overall brightness variations were factors of  6.2, 5.8, 6.5, 3.9, 3.2, and 3.0 at V, B, U, UVW1, UVM2 and UVW2, respectively. Figure \ref{fig4} also shows that the flare maximum occurred on MJD 55518-20, although the peaks at the various bands were not all simultaneous. This was followed by a gradual decline interspersed by two smaller flares.

\begin{figure}
\epsscale{0.9}
\plotone{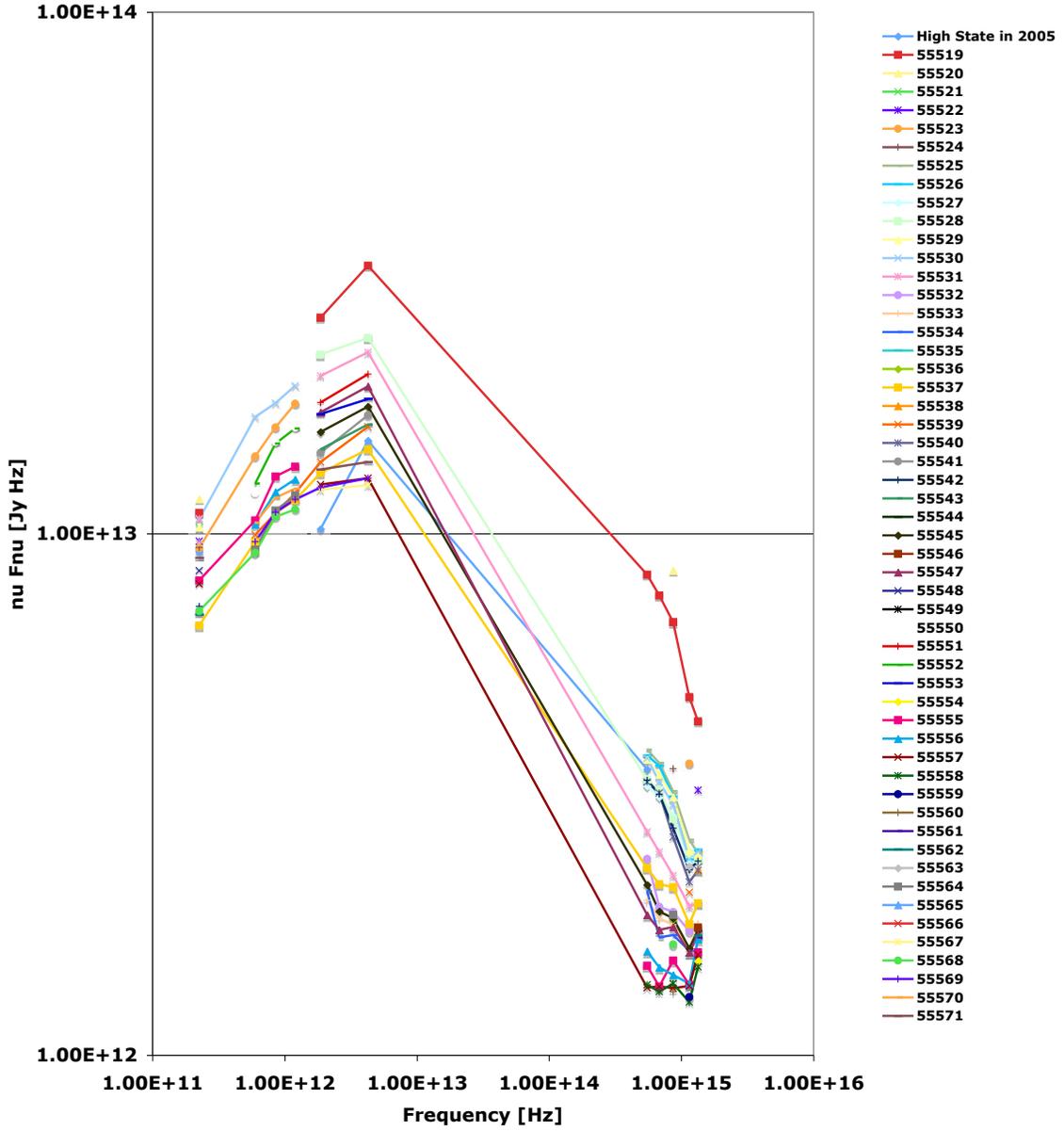}
\caption{Spectral energy distributions from millimeter through ultraviolet bands.  The symbols and colors represent different MJDs.
\label{fig2}}
\end{figure}

\begin{figure}
\epsscale{0.4}
\plotone{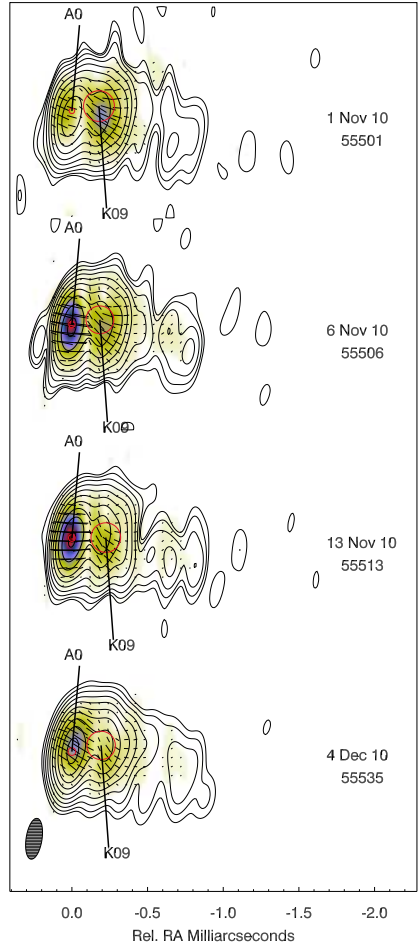}
\caption{43 GHz VLBA images of 3C~454.3 in November-December 2010, convolved with an elliptical restoring beam of FWHM
dimensions $0.33\times0.14$ mas oriented along position angle $-10^\circ$. Contours represent total intensity with levels
in factors of 2 from 0.15\% to 76.8\% of the global peak brightness of 18.47 Jy beam$^{-1}$. Colors correspond
to linearly polarized intensity with a global peak of 0.71 Jy beam$^{-1}$ (red).
The short bars indicate the relative amplitude and direction of the polarization electric vector.
The core (A0) and knot K09 are indicated with red circles delineating the FWHM size of the Gaussian model components.
New knot K10 is blended with A0 on these images. The flux density of the core is
10.3, 14.1, 14.2, and 17.7 Jy (uncertainty $\approx 10\%$) from the first to the last image displayed.
See \citet{jorstad2012} for details.
\label{fig3}}
\end{figure}

\begin{figure}
\epsscale{1.0}
\plotone{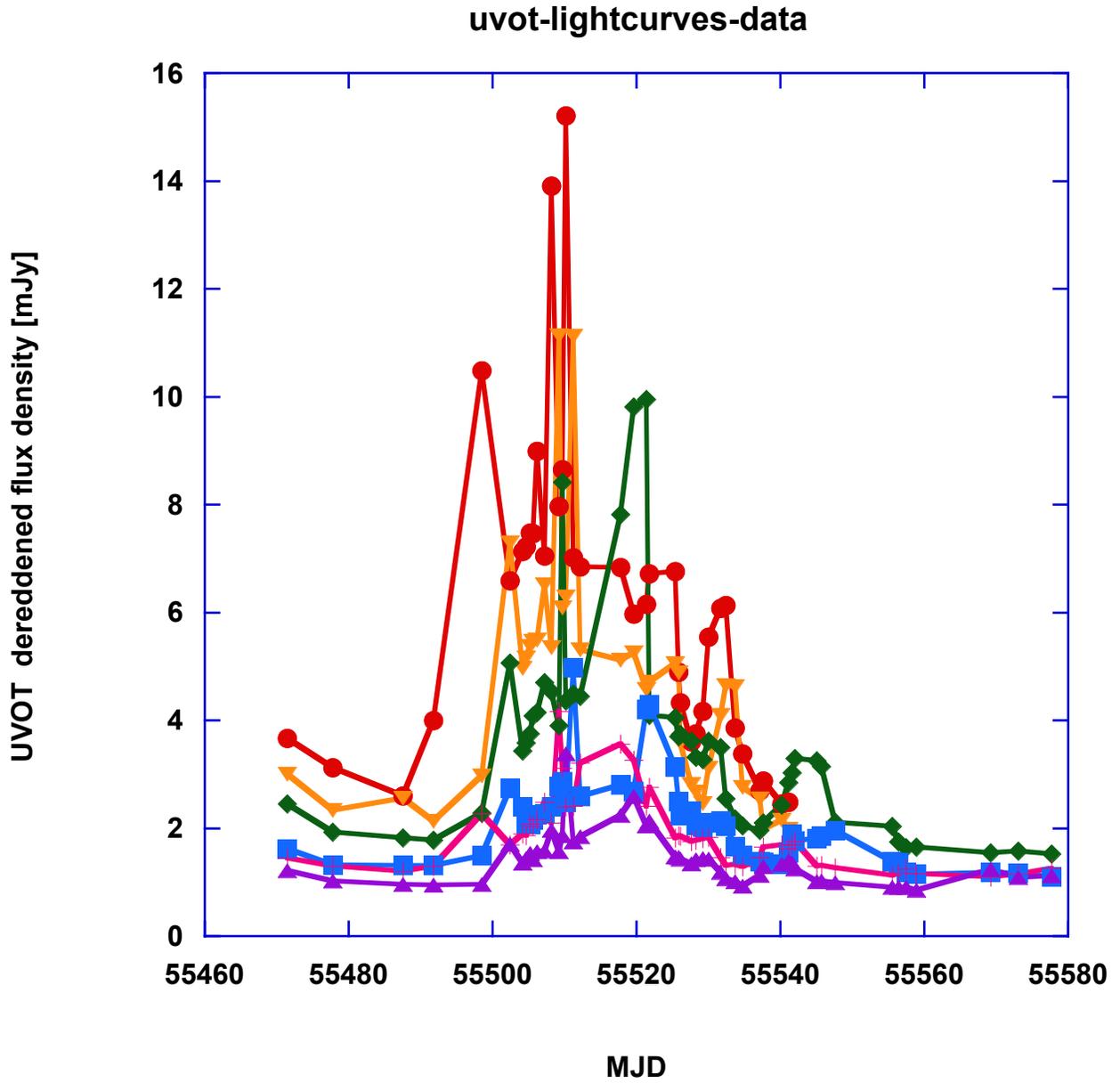}
\caption{Swift UVOT lightcurves before, during and after flare.  Filter bands include V (red circles), B (orange
inverted triangles), U (green diamonds), UVW1 (blue squares), UVM2 (pink-red crosses), and
UVW2 (purple triangles). Dereddening has been applied as described in Raiteri et al. 2011. \label{fig4}}
\end{figure}

The data in the six UVOT bands cannot be fit by a simple power-law because of features possibly caused by emission lines (e.g., a noticeable bump in the UVM2 spectrum caused by Ly$\alpha$, with smaller contributions by CIV in UVW1 and OVI+Ly$\beta$ in UVW2) and the Galactic interstellar dust absorption band at 2175~\AA, as described by \citet{raiteri2011} and references therein. These features are clearly visible in the overall millimeter-infrared-optical-ultraviolet SEDs in Figure \ref{fig2}. The spectral index using only V and UVW2 bands (ignoring the bumps and dips in between) can be readily calculated: it is much steeper at the flare peak than near the end of the {\it Herschel} observations --- $1.70 \pm 0.02$ on MJD 55519 and $1.00 \pm 0.01$ on MJD 55558. 

\begin{figure}
\plotone{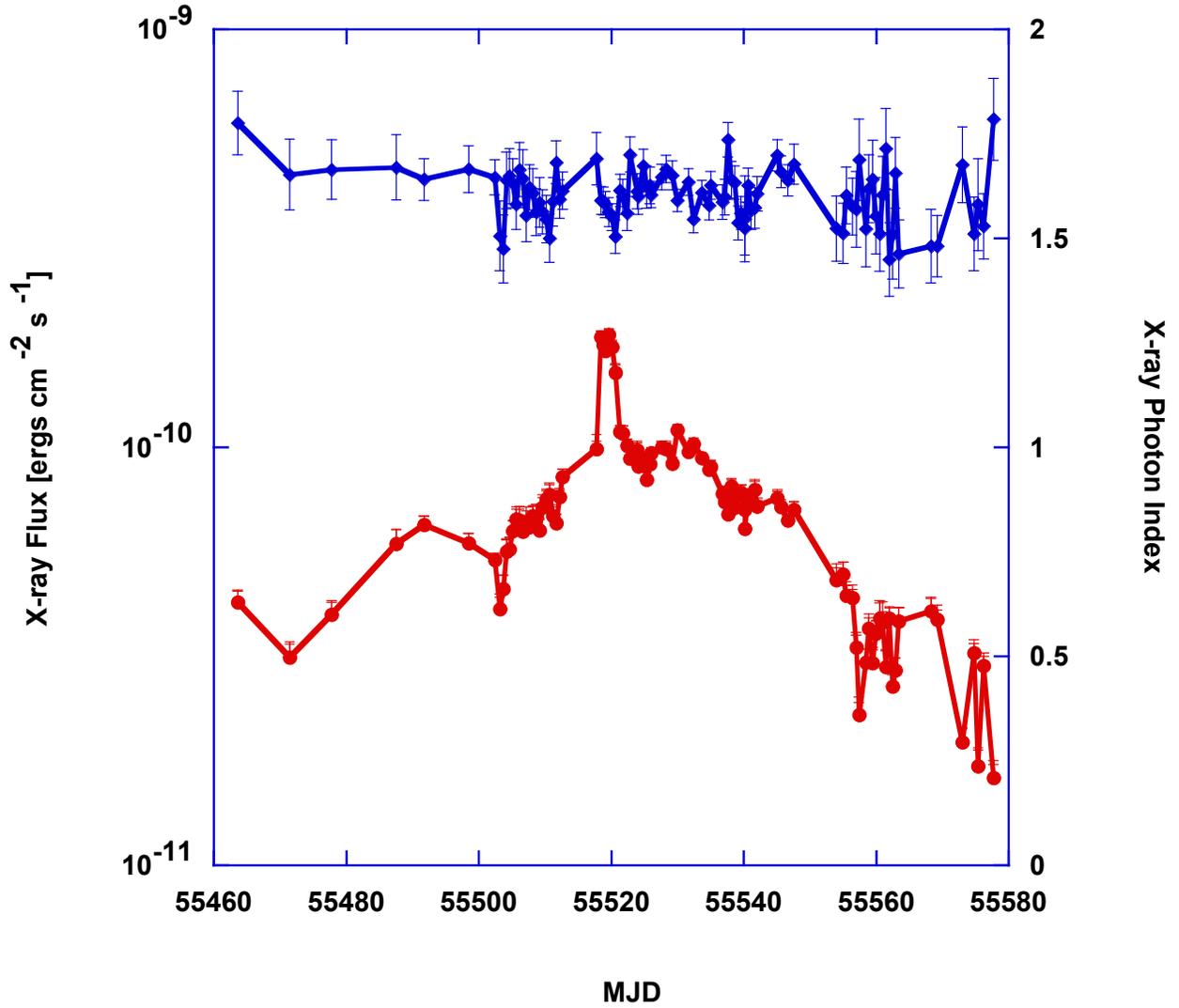}
\caption{X-ray light curve (red circles) and photon index (blue diamonds)
before, during and after the flare. The X-ray photon index varied between $\sim$ 1.5-1.7 and became slightly
harder during the 2-3 days of the flare peak. 
\label{fig5}}
\end{figure}

\begin{figure}
\plotone{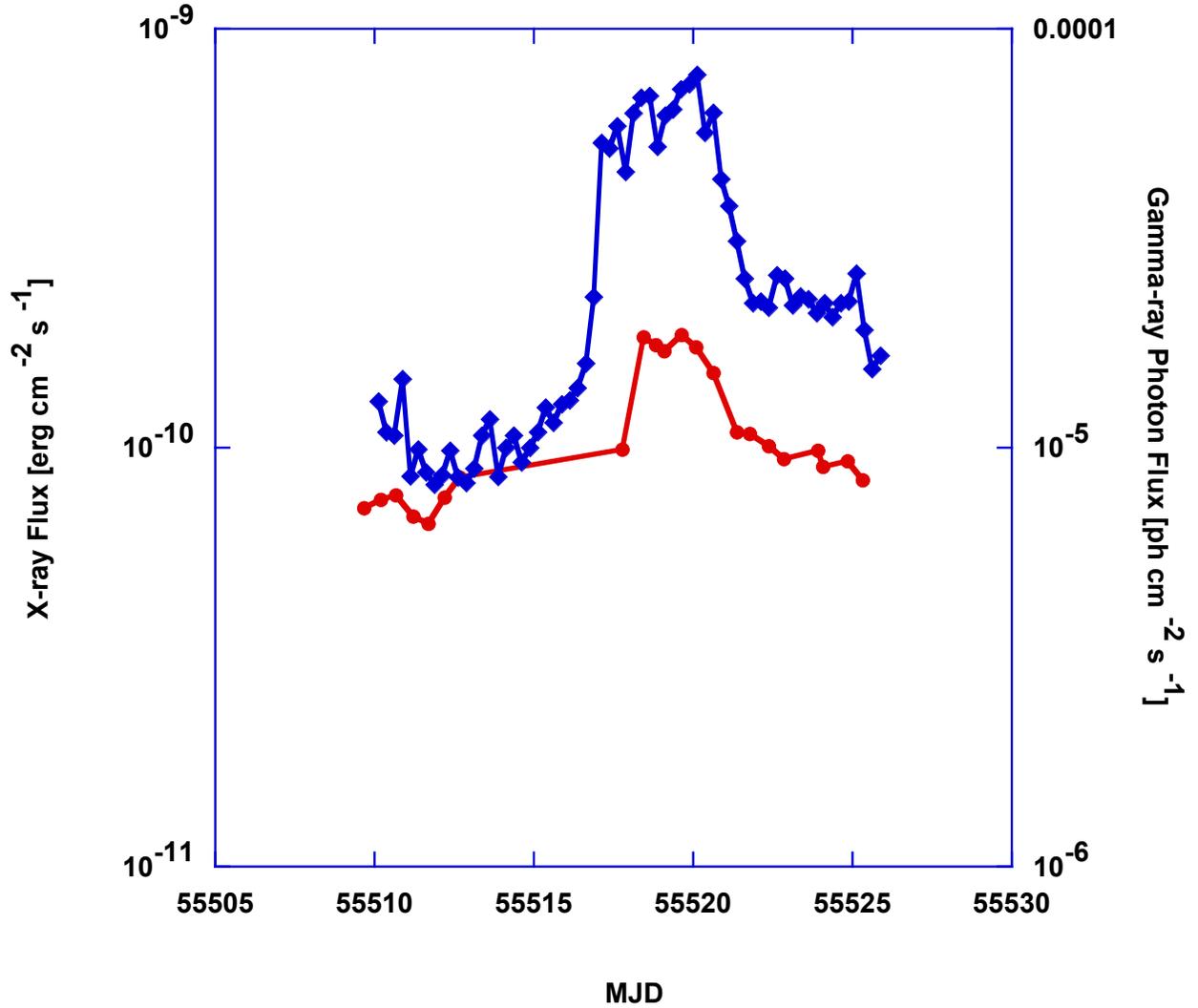}
\caption{X-ray and $\gamma-$ray light curves centered on date of the main flare. X-ray flux (red circles) and 6-hour binned $\gamma-$ray photon flux (blue diamonds) show general agreement with differing details, especially during the peak of the flare. 
\label{fig6}}
\end{figure}

The {\it Swift} X-ray light curve and photon spectral indices are presented in Figure \ref{fig5}. 
Figure \ref{fig6} shows the X-ray and 6-hour-binned $\gamma$-ray light curves for fifteen days (MJD 55510-55525) encompassing the brightest days of the flare. The peak in the X-ray flare extended over five days between MJD 55517.7762 and MJD 55522.8419, with two nearly identical maxima of $1.84\times 10^{-10} \pm 5.9\times 10^{-12}$ erg cm$^{-2}$ s$^{-1}$ and $1.86\times 10^{-10} \pm 6.7\times 10^{-12}$ erg cm$^{-2}$ s$^{-1}$ on MJD 55517.7762 and MJD 55519.6305, respectively. During the brightest days of the flare on MJD 55519-20, the X-rays increased by a factor of 2.8, and the $\gamma$-rays by 8.9, over the low values on MJD 55511. The most substantial variation was observed between MJD 55517.7762 and 55518.4373, when the X-ray flux rose from $9.91\times 10^{-11}$ erg cm$^{-2}$ s$^{-1}$ to $1.84\times 10^{-10}$ erg cm$^{-2}$ s$^{-1}$, an increase of 86\% in 16 hours. In contrast, \citet{abdo2011} reported that the fastest $\gamma$-ray flux change was on MJD 55516.5, when the flux increased by a factor of 4 in 12 hours, for a doubling time of 6 hours. We have no Swift data between MJD 55512.6725 and 55517.7762, hence we cannot tell whether the X-ray and $\gamma$-ray flares rose exactly simultaneously on MJD 55516-55517.   

During the {\it Swift} observations of the flare between MJD 55502 and 55577, the photon index increased from 1.45 to 1.79, with an average of $1.60\pm 0.06$.
In the April 2008-March 2010 period, \citet{raiteri2011} found that the photon index varied from 1.38 to 1.85, with an average value of 1.59, nearly the same as our value in November 2010-January 2011. During MJD 55502-55577, the X-ray flux varied from $1.62 \times 10^{-11} \pm 1.4 \times 10^{-12}$ ph~cm$^{-2}$~s$^{-1}$ to $1.86\times 10^{-10} \pm 6.2\times 10^{-12}$ ph~cm$^{-2}$~s$^{-1}$ (the latter on MJD 55519.6305), averaging $3.78 \times 10^{-11}$ ph~cm$^{-2}$~s$^{-1}$. There was no clear tendency for the photon index to flatten during the highest points of the flare. During the entire flare period, 3C~454.3 was much brighter than usual, and it did not return to very low, quiescent levels during the course of our observations. 

\subsubsection {Fermi LAT}

\begin{figure}
\plotone{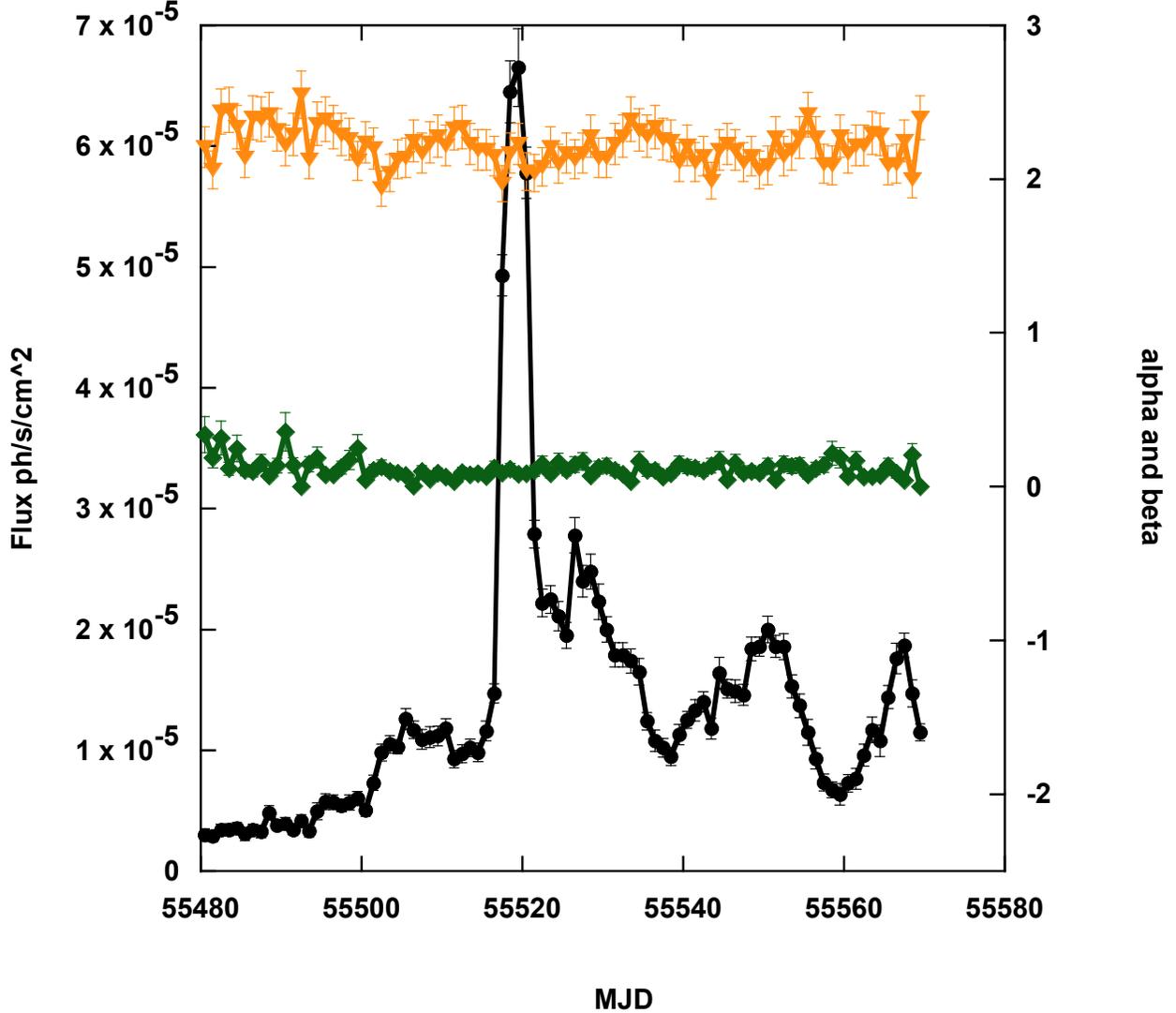}
\caption{Fermi LAT $\gamma-$ray light curve made with 1-day bins. {\bf The photon flux is shown as black filled circles. The parameters of the log-parabolic fit, $\alpha$ and $\beta$, are shown as orange triangles and green diamonds, respectively. The errors on the photon flux are as shown in Table 6; we plot 6\% errors on $\alpha$ and 35\% errors on $\beta$. See text for discussion of correlated errors.}
\label{fig7}}
\end{figure}

The $\gamma$-ray light curve based on 1-day binned data is displayed in Figure \ref{fig7}. The main flare peak around MJD 55519-20 was followed by a rapid decline with three smaller peaks occurring on MJDs 55526, 55550, and 55567. The minimum during the interval of the {\it Herschel} observations occurred on MJD 55558-9. Overall, the flux varied by a factor of 10.5 from MJD 55519.5 to 55559.5. As described by \citet{abdo2011},
$\gamma$-ray spectral variations were modest, although the spectrum flattened somewhat near the
peak of the main flare. Within the log-parabolic spectral model that we have employed, this implies that the peak of the
$\gamma$-ray SED shifted to somewhat higher energy as the flux peaked.

\subsection {Correlations and Time Delays between Bands}

\begin{figure}
\plottwo{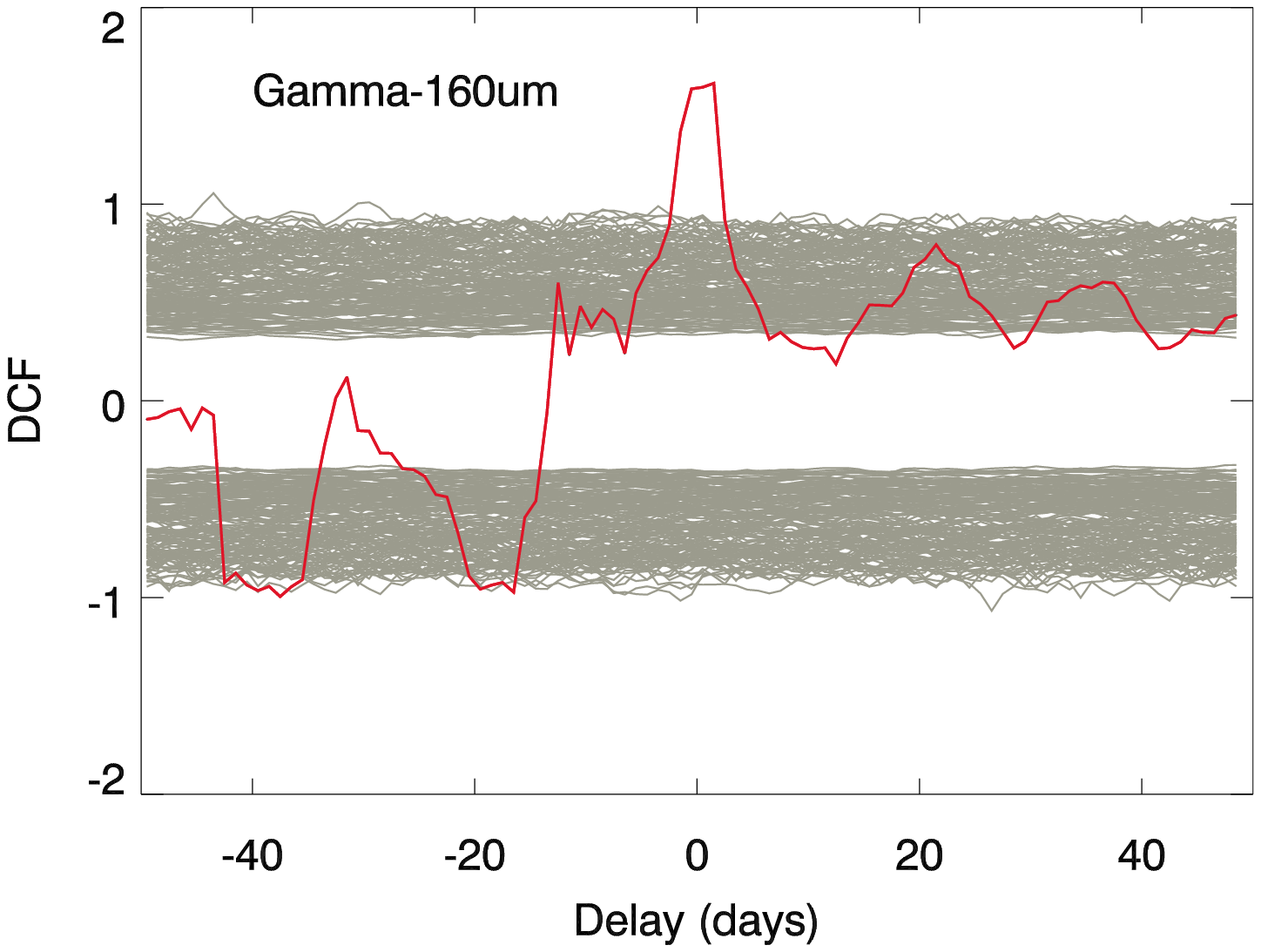}{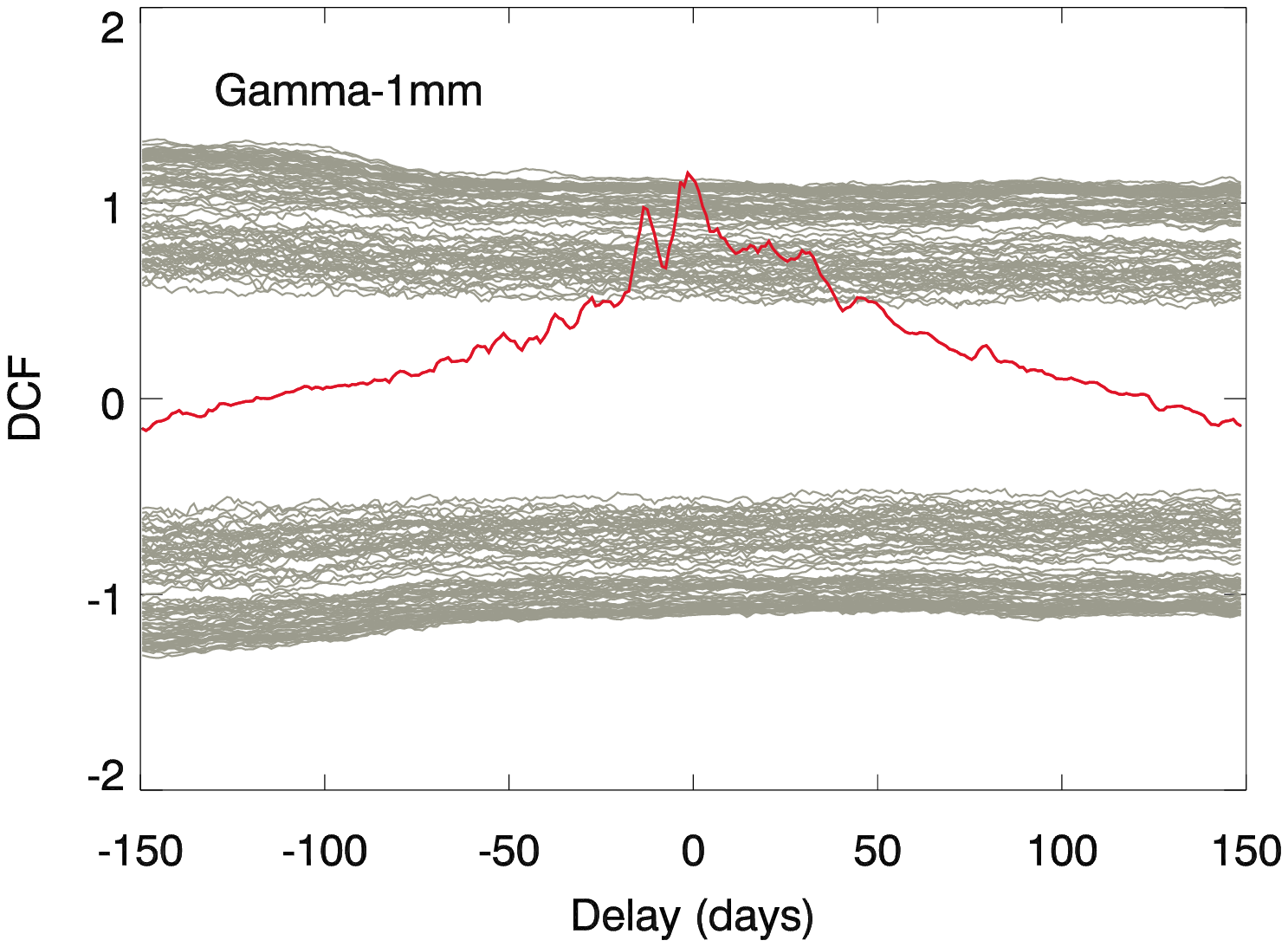}
\caption{Discrete correlation function (red curve) for {\it (Left)} $\gamma-$ray and 160 $\mu$m data and
{\it (right)} $\gamma-$ray and 1.3 mm data. Gray curves correspond to results of Monte Carlo simulations (see text). In both cases, the peak in the DCF has been fit with a Gaussian, from which we measure a delay of
$-0.5$ to $+1.5$ days
between $\gamma-$ray and 160 $\mu$m variations and $-1.5$ to +3.5 days between $\gamma-$ray and 1.3 mm variations. 
\label{fig8}}
\end{figure}

\begin{figure}
\plotone{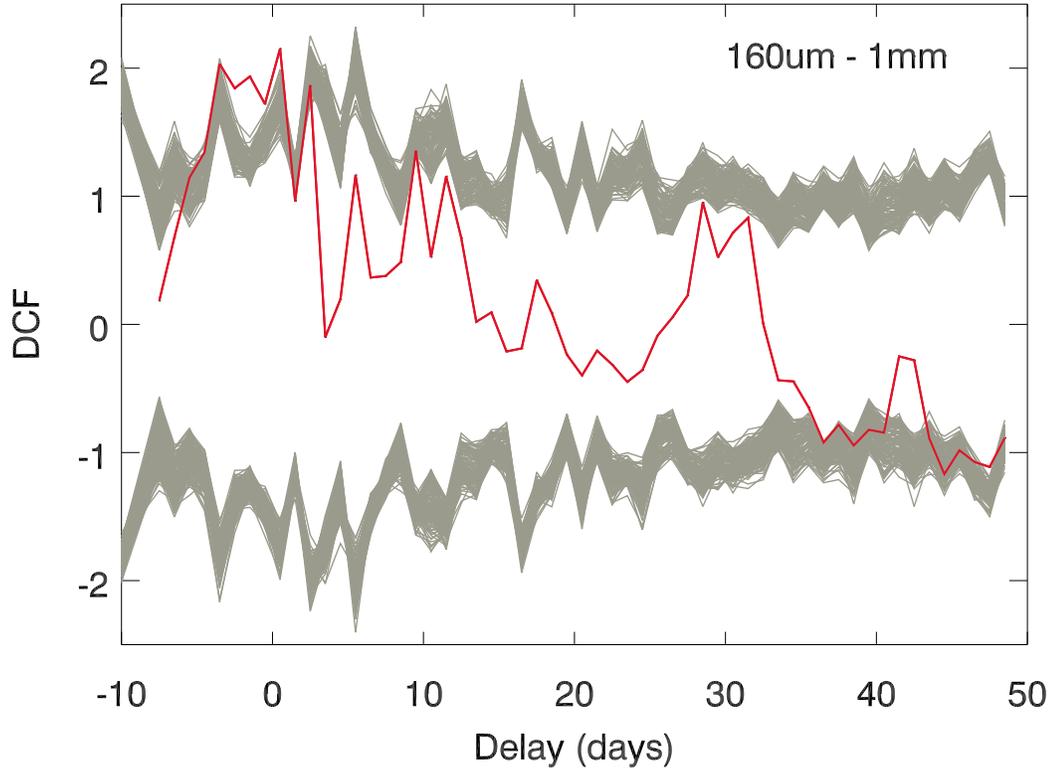}
\caption{Discrete correlation function (DCF, red curve) for 160 $\mu$m and 1.3 mm data. Gray curves correspond to results of Monte Carlo simulations (see text). The DCF is not plotted if there are too few data points to calculate it at
a given time delay, hence the red curve is discontinuous.
The peak in the DCF has been fit with a Gaussian from which we measure a delay between $-3.5$ and $+0.5$ days.
\label{fig9}}
\end{figure}

We evaluate the time delays between bands by applying the discrete correlation function (DCF) methodology \citep{edelson1988}, as implemented in the {\it aitlib} library in IDL.  In summary, the results of the DCF analysis indicate that variations at 0.1-200 GeV $\gamma$-ray energies, 160~$\mu$m,
and 1.3~mm were simultaneous to within the accuracy of the method. For the correlation analysis, we supplement the 160 $\mu$m data with 250 $\mu$m data scaled by 0.724 (the average of the ratio of 250 $\mu$m to 160 $\mu$m flux densities on days when both bands were observed with different instruments on {\it Herschel}), resulting in 25 points in the light curve. The 1.3 mm SMA data over the time range from 2008 January 12 to 2011 October 27 are used with the original sampling (350 points). The {\it Fermi} LAT $\gamma$-ray light curve corresponds to one-day binned data from the beginning of science operations
on 2008 August 5 to 2011 October 21 (1167 points). The resulting DCF curves are shown in Figures \ref{fig8} and \ref{fig9}.

In order to determine the 3-$\sigma$ level of significance of a given correlation, we follow the
methodology proposed by \citet{chatterjee2008}, \citet{max-moerbeck2010} and \citet{agudo2011a}. We simulate 5000 light curves for each
of the three wavebands based on the mean, standard deviation, and slope $b$ of the power spectral density (PSD) of
the observed flux variations. The PSD, which corresponds to the power in the variations as a function of timescale,
is fit by a power law, $P(\tau) \propto \tau^{b}$, where $\tau$ is the time-scale of the variations. [According to
\citet{abdo2010}, the PSDs of $\gamma$-ray light curves of quasars have average slopes of $\langle b \rangle = 1.7$.]
In the simulations, we vary the value of $b$ from 1.0 to 2.5 in increments of 0.1. The simulated light curves,
binned in time in exactly the same way as the actual light curves, were cross-correlated to determine the
probability of obtaining via random chance a particular value of the DCF at each relative time lag. This
allows us to determine, for each pair of $b$ values, the level at which the measured DCF is significant with
99.7\% confidence (3-$\sigma$), thus testing whether the observed correlation (or anti-correlation) is
statistically significant. Curves representing these 99.7\% confidence levels are drawn in gray in
Figures \ref{fig8} and \ref{fig9}.

Figures \ref{fig8} and \ref{fig9} show that (1) the correlation between the $\gamma$-ray and 160 $\mu$m light curves is significant for delays from -0.5 to +1.5 days (independent of the value of $b$), where a negative delay corresponds to $\gamma$-ray leading IR variations; (2) the correlation between the $\gamma$-ray and 1.3~mm light curves is significant and independent of the PSD slope for delays from -1.5 to +3.5 days, although there is a second DCF peak at -13.5$\pm$1~days that is significant for $b<$2.0 for both light curves;
(3) the correlation between the 160 $\mu$m and 1.3~mm light curves is significant for delays from -3.5 to +0.5 days, where a negative delay corresponds to the IR variations leading. 

The time delays at the peaks of the DCFs all include zero, but the offset from zero of the central values of 
the delays indicates that, on average, significant time lags occur. These are in the sense that the
$\gamma$-ray variations lag those at 160 $\mu$m by $1\pm 0.5$ days, while the 1.3 mm variations lag those
at 160 $\mu$m by $1.5\pm 2$ days. We therefore conclude that, to within the accuracy of our analysis, the
variations are simultaneous at all three wavebands. Inspection of the profiles of the main flare confirm that
the time lags were essentially zero for this event.

\subsection{Relationship between Millimeter and Submillimeter Variations and Features on the VLBA Images}

\begin{figure}
\plotone{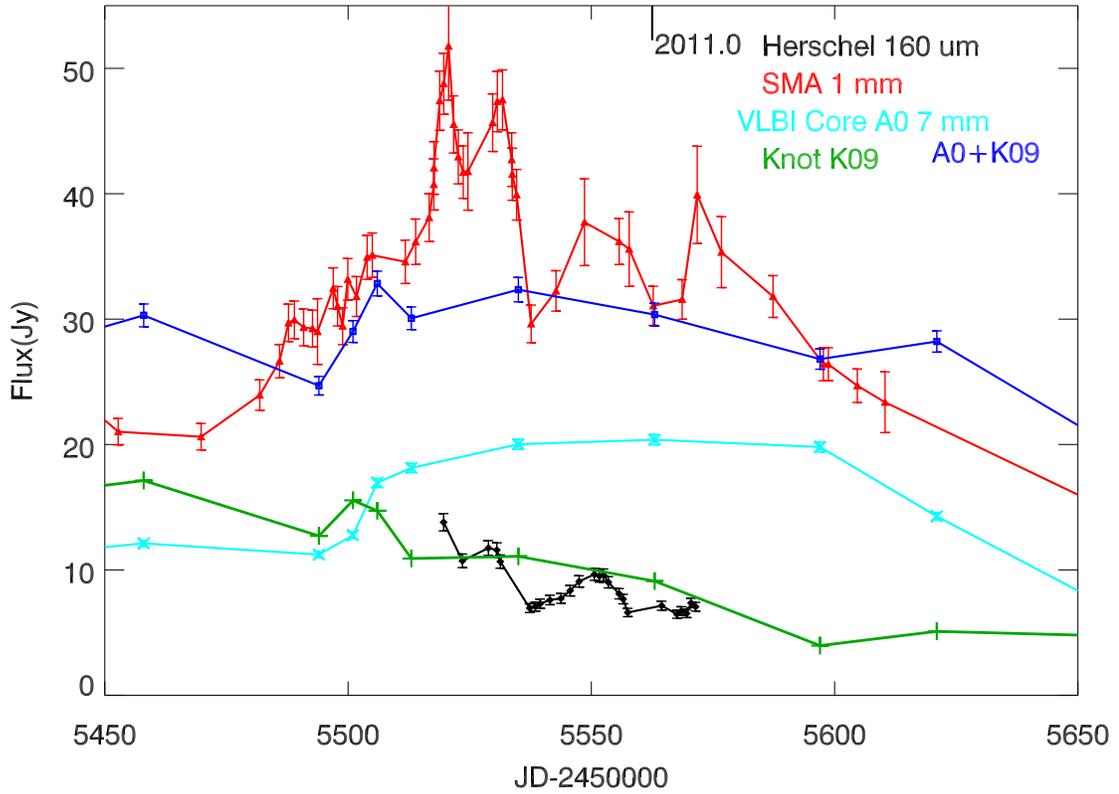}
\caption{Light curves of total flux density observed at 160 $\mu$m (black diamonds; including interpolated fluxes based on 250 $\mu$m data), and 1.3 mm (red triangles), individual 43 GHz flux densities of the core A0 (purple Xs) and knot K09 (green crosses), and the sum of 43 GHz flux densities in the core A0 and Knot K09 (magneta squares and dotted line).
\label{fig10}}
\end{figure}

The sequence of VLBA images displayed in Figure \ref{fig3} reveal two main emission features within 0.3 mas
(2 pc projected distance) of the upstream end of the jet, A0 (the ``core'' plus new superluminal knot
K10) and superluminal knot K09 that passed through the core in late 2009 \citep{jorstad2012}. The images
reveal a sharp increase in the brightness of the core during the first five weeks of the outburst. 
Figure \ref{fig10} presents the 1.3 mm and 160 $\mu$m (including scaled 250 $\mu$m data) light curves along
with that of A0 and K09 at 7 mm. If they are responsible for
the 1.3 mm and far-IR emission, then A0 and/or K09 should vary synchronously with the SMA 1.3 mm and
{\it Herschel} 160 $\mu$m fluxes. Indeed,
the overall trend of the SMA flux matches that of A0 during the period MJD=55500 to 55580.
The flux of K09, on the other hand, underwent a downward trend during the period shown, from
about 16 Jy to 4 Jy.  The nearly flat flux density spectrum between 225 GHz and 43 GHz (A0 + K09 only)
is slightly inverted during the flare ($\alpha=-0.12$ to $-0.05$).
Knot K09 must have possessed a steep spectral index, since the sum of the A0 and K09 fluxes exceeded the
1.3~mm flux before MJD 55500 when K09 was brighter than the core. The 7~mm A0 and 1.3~mm fluxes varied
rapidly from MJD=55500 to 55515, on a time-scale much shorter than a month. Based on the overall behavior
of the light curves, it is highly likely that the outburst at 1.3~mm
is connected with the outburst in A0, although the data are ambiguous
because the VLBA measurements happened to occur near or at local minima in the SMA light curve.

\section{Spectral Energy Distributions}

The combination of several satellite observatories and a number of ground-based telescopes has
provided an extraordinarily rich dataset. The multi-waveband light curves provide both guidance
and challenges for the development of theoretical models. 

\subsubsection{Overall Characteristics of 52 SEDs}

\begin{figure}
\plotone{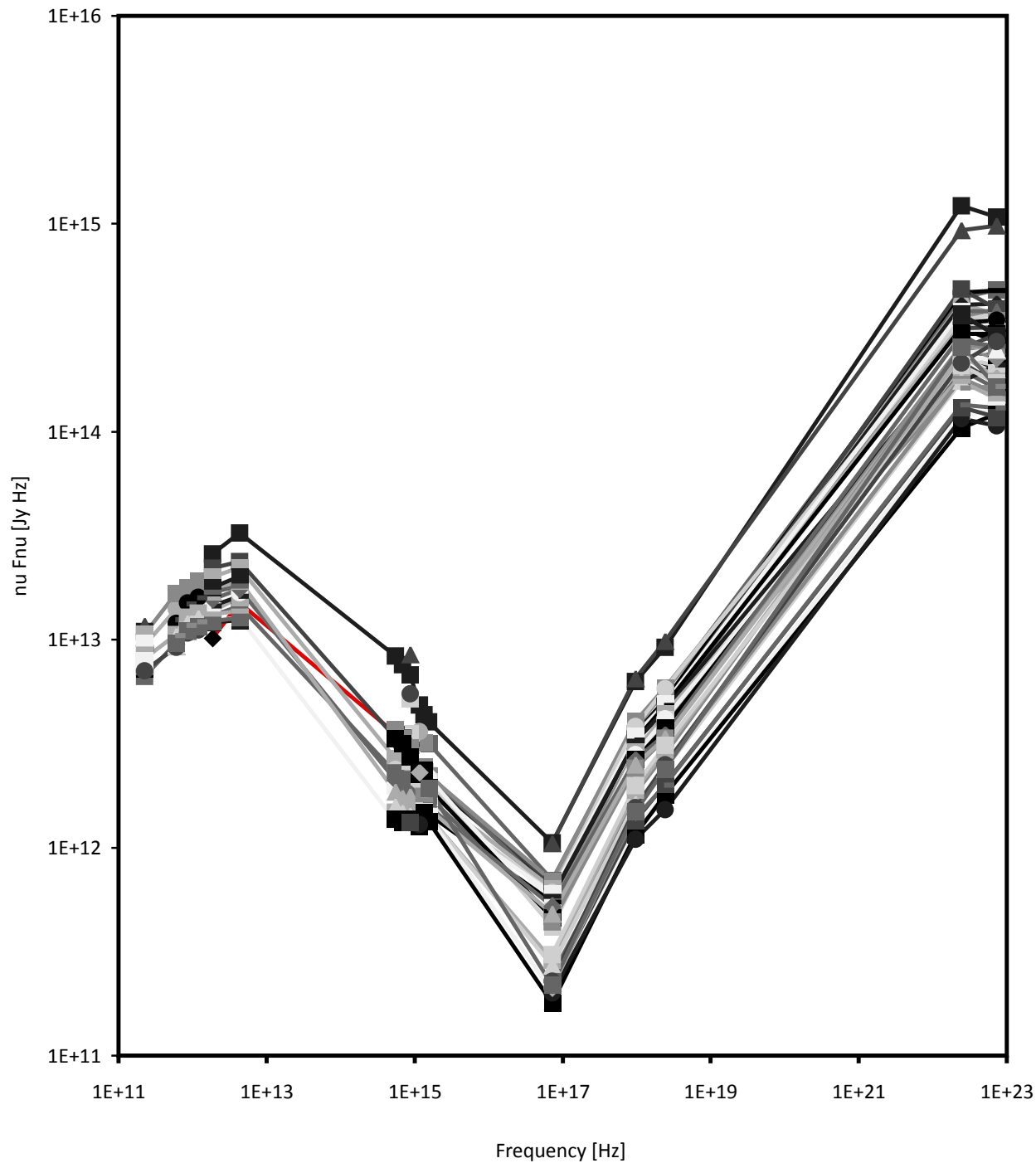}
\caption{Spectral energy distributions obtained on 52 days during Nov 2010-Jan 2011 (MJD 55519-55571)and from the high state in 2005 (in red line visible from $\sim 10^{12}$ to $\sim 10^{15}$Hz ), the latter from Ogle et al. 2011. The gray-to-black shaded symbols and lines represent different dates. Note the similarity in overall appearance and that the 2005 high state SED blends in with the other SEDs. 
\label{fig11}}
\end{figure}

We form 52 SEDs over the time interval of the near-daily observations, from 2010 November 19 to 2011 January 10 (see Tables 7 and 8). The SEDs are remarkably similar to each other, differentiated mainly by small day-to-day fluctuations (Figure \ref{fig11}). The peak amplitude of the synchrotron emission varied by a factor of $\sim 2$, while the peak amplitude of the inverse Compton emission varied by a factor of $\sim 10$. The millimeter to far-IR spectral indices (which vary only slightly) are very similar to the X-ray spectral indices (which are also essentially constant).
The mid-IR SED, measured with the IRTF on 2010 November 3 near the start of the outburst, peaked at $\sim 15$ $\mu$m.
We cannot determine whether the peak in the IR SED shifted upward in frequency during the flare as it did in 2005 \citep{ogle2011}, since the maximum in the SED is not clearly defined by our {\it Herschel} far-IR observations.
The peak in the $\gamma$-ray SED changed only modestly during the flare \citep[see Fig. \ref{fig11} and][]{abdo2011},
although the low-energy cut-off to the $\gamma$-ray spectrum of 100 MeV reduces our ability to define
the maximum when it is below $\sim 500$ MeV.
Improvements in the $\it Fermi$ LAT data processing should allow extension of the measured
spectra to lower energies in the future. This will provide further constraints on emission models.

Both during and after the peak of the flare, the difference between the submillimeter and optical 
spectral slopes exceeded the value of 0.5 which is expected from synchrotron and 
inverse Compton radiation losses. Specific details are as follows. Between the 1.3 mm and 160 $\mu$m bands, the spectral index $\alpha$ (where $S_\nu \propto \nu^{-\alpha}$) was $0.59 \pm 0.02 , 0.72 \pm 0.03, 0.69 \pm 0.03$ and $0.83 \pm 0.03$ on MJDs 55519, 55531, 55537 and 55571, respectively. Between the 70 $\mu$m and Swift V band, the value of $\alpha$ was $1.27 \pm 0.03, 1.57 \pm 0.04$, and $1.51 \pm 0.04$ on MJDs 55519, 55531 and 55537, respectively (no Swift V band data were obtained on MJD 55571).
The breaks in the synchrotron spectrum correspond to changes in slope of 0.68, 0.85, and 0.82. 

The millimeter-infrared SEDs are shown in Figure \ref{fig2}. We note that at 350 $\mu$m, the SEDs showed an inflection point where the SED is higher than it would be with a linear interpolation between 500 $\mu$m and 250 $\mu$m. This small bump in the synchrotron peak during the declining phase of the outburst may be due to the superposition of two (or more) synchrotron-emitting components, or simply to inhomogeneities across a single emission feature.  Given the steep spectrum of knot K09, its flux at 350 $\mu$m is far too low to cause the small bump at 350 $\mu$m. In 2005, the millimeter- infrared SED, observed with the SMA and {\it Spitzer}, contained a dip at 160 $\mu$m \citep{ogle2011}, which was interpreted as the valley between two synchrotron peaks, one of which occurred between 850 $\mu$m and 160 $\mu$m where there was no observational data.  We note that 3C~279 showed the same type of SED dip in {\it Spitzer} observations \citep{abdo2012}. The dip is unlikely to be caused by an instrumental calibration problem, since it has been observed in two blazars by three different spacecraft (ISO, {\it Spitzer}, and {\it Herschel}) with independent instrument calibration strategies. Multiple or inhomogeneous emission regions therefore appear
to be common features in the synchrotron SEDs of blazars.

The one-day binned Fermi LAT data were well-fit by a log parabolic model on most days, with one day on which the data were better fit by a power law. The amplitude of the $\gamma$-ray SED peak varied by a factor of $\sim$ 10 between the flare on MJD 55519 and the low, post-flare level of MJD 55558.
The break energy was typically 300 MeV, with an increase to 345 MeV at the peak of the main flare.  

The {\it Herschel} infrared observations began, coincidentally, on the day on which the flare peaked. 
The above correlation analysis of the far-IR and $\gamma$-ray light curves demonstrates that the variations were simultaneous, with any delays being less than about 1.5 days. While the very short time delays
across frequencies indicate that a coherent emission zone is responsible for the outburst,
the differences in the detailed flare profiles imply that there are irregularities in the
physical conditions within that zone. These can take the form of gradients or spatial
irregularities in the relativistic electron density and energy distribution or magnetic field
strength and direction, for example.
 
The SMA 1.3 mm and 160-250 $\mu$m variations (Fig. \ref{fig10}) are also essentially simultaneous, with any delays less than 3.5 days. This indicates that the flaring synchrotron emission was optically thin at millimeter bands, and was probably dominated by a single variable component in late 2010. This contrasts with the
conclusion of \citet{ogle2011} that, several months after a major outburst in 2005, the emission
arose from two components within a twisted jet. It also falsifies the model proposed by
\citet{vercellone2011} that places the site of the flare within a plasmoid of radius $3.6\times 10^{16}$
cm at a distance 0.05 pc from the center of the accretion disk. Besides locating the flare at a much
smaller distance from the central engine than indicated by the VLBA observations, that model adopts
physical parameters corresponding to an optical depth to synchrotron self-absorption of 5600
at 1.3 mm.

Examination of the light curves reveals that 3C~454.3 brightened significantly at both 1.3 mm and $\gamma$-ray energies weeks before the $\gamma$-ray brightness flared precipitously. This implies that the disturbance in the jet flow intensified gradually, starting well before the main flare began. The entire outburst therefore
appears to be a prolonged event rather than an impulsive surge of energy, while the main flare does
appear to be rather impulsive. This could be related to the complex structure (with three
essentially stationary brightness peaks) of the core region inferred from analysis of VLBA images by
\citet{J10}. The passage of a disturbance in the jet flow across each stationary feature in the
jet would be expected to cause an outburst with complex temporal structure.
\section{Theoretical Modeling of the SED}
\subsection{General Considerations}

There are a number of features of the SED of 3C~454.3 during the outburst that any successful model
should be able to reproduce. One is that the two SED peaks do not move sharply toward lower
frequencies as the flux declines, as expected in expanding plasma blob or shock models \citep[e.g.,][]{mg85}.
The strong steepening of the spectra above the SED peaks from radiative energy losses
after impulsive injection of relativistic electrons is also absent. Perhaps the most important
aspect of the outburst is the general correspondence of the main flare at submillimeter to $\gamma$-ray
wavelengths combined with significant dichotomy in flare substructure, as well as different amplitudes
of the secondary flares, at the various wavebands. 
The similarity of the far-IR and X-ray spectral indices conforms with the expectations of the
standard emission model for quasars in the blazar class, in which the radio to UV nonthermal emission is
synchrotron radiation and the X-ray to $\gamma$-ray emission is inverse Compton scattering, with the
same population of relativistic electrons responsible for both. We therefore interpret the data in
terms of this scenario.

Previous interpretations of outbursts in $\gamma$-ray bright quasars concentrate on models in which
the seed photons are from the accretion disk, BLR, or dust torus
\citep[e.g.,][]{bla00,sikora09,tav10,vercellone2011}. Such sources
of seed photons from outside the jet can produce very high $\gamma$-ray fluxes in the observer's frame
owing to extra Doppler boosting of the seed photons in the plasma frame, as long as the emitting plasma
is not well downstream of the source of seed photons.
In these models, however, the seed photon field varies smoothly
as the emitting plasma propagates down the jet. In contrast, the flares have jagged time profiles in
3C~454.3 and other blazars, and $\gamma$-ray and optical variations do not correspond well in detail
\citep{bonning09,raiteri2011,vercellone2011, agudo2011b}.
In models in which the magnetic field is either uniform or completely chaotic,
rapid fluctuations of the $\gamma$-ray flux would need to be caused by
variations in the relativistic electron distribution, but these should affect the optical emission in
the same way \citep[see][for a discussion]{mar10}. A possible exception would be a model 
incorporating strong fluctuations in the velocity structure of the flow so that the Doppler factor
varies across the emission region \citep{gian09,np12}.

The differences in the light curves can be caused by spatial and temporal variations in the direction
of the magnetic field and/or acceleration of electrons. The former, which can be caused by turbulence in
the jet flow, agrees with the polarization of the synchrotron emission. Time-variable degrees of polarization
from a few to 30\%, as observed at optical wavelengths in blazars --- including 3C~454.3 in 2010 November
\citep{jorstad2012} --- can be reproduced by a model consisting of 10--100 cells of emission, each
with a uniform magnetic field with random orientation relative to the other cells \citep{mj10}.

The lack of strong evolution of the outburst toward lower frequencies favors emission by a stationary
structure such as a standing shock rather than by a self-contained shock or plasmoid that expands as
it moves down the jet. This leads us to adopt the scenario of \citet{mj10} in which turbulent plasma
in the jet crosses a standing shock with a conical geometry \citep[see also][]{caw06}.
Such a model is capable of producing different flare profiles at optical and $\gamma$-ray frequencies
because the changing magnetic field direction strongly affects the synchrotron flux without much, if any,
influence on the inverse Compton emission \citep{mar12}. Further differences at the two wavebands
can occur if there are strong velocity fluctuations in the flow that
create spatial variations in the Doppler factor, or if the main source of seed photons is local to the jet,
not spread out as would be the case for photons from the BLR or dusty torus.

\subsection{Model Involving Turbulent Plasma Flowing across a Stationary Conical Shock Plus Mach Disk}

\begin{figure}
\plotone{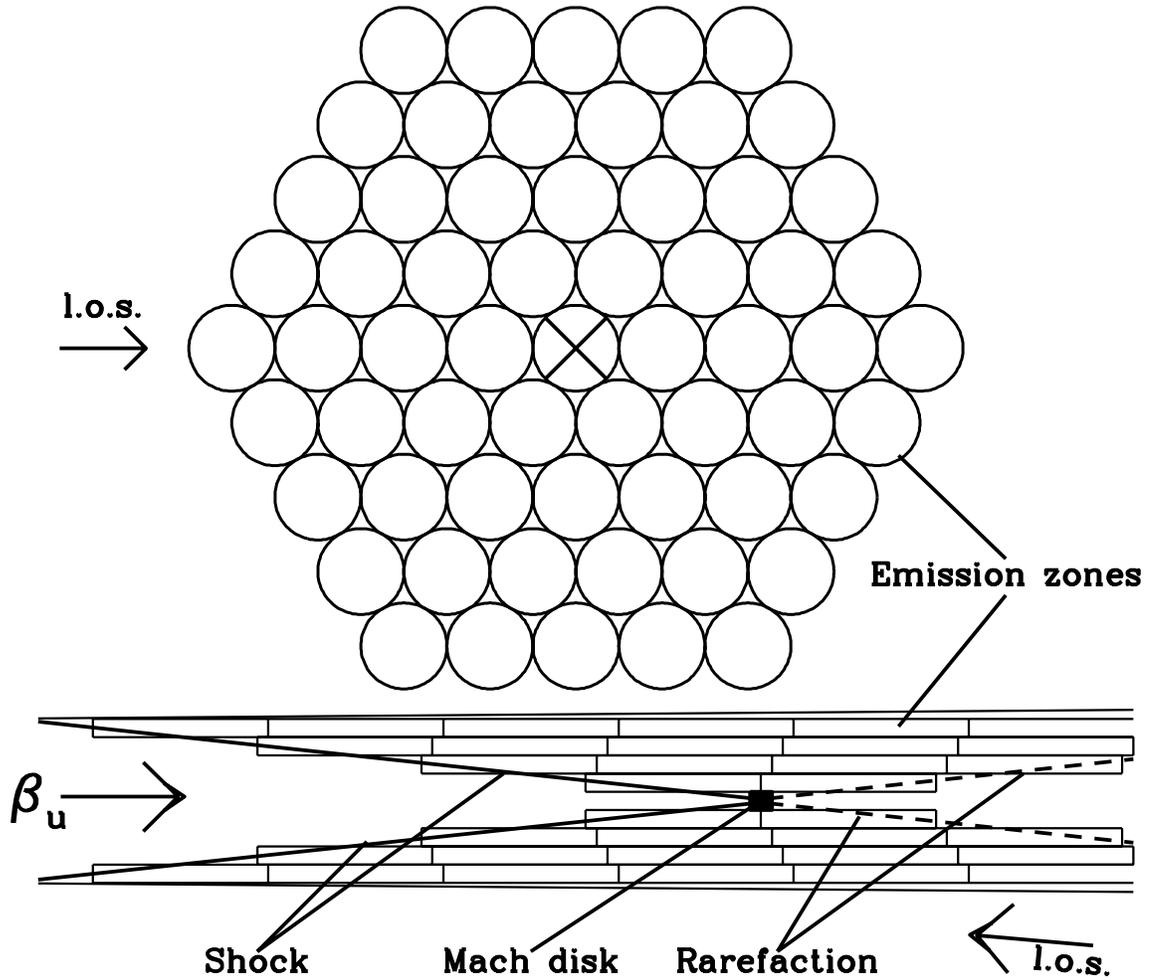}
\caption{Sketch of geometry adopted in the TEMZ model. {\it Top:} View down
the jet, whose cross-section is divided into many cylindrical cells. Actual
calculations displayed in Fig. \ref{fig13} use 271 cells across the jet.
{\it Bottom:} Side-on view. Cylindrical cells
appear as rectangles. The conical shock compresses the flow \& accelerates electrons.
The Mach disk, if present, is at the axis, oriented transverse to the flow. 
\label{fig12}}
\end{figure}

In order to determine whether a model involving turbulent jet plasma flowing across a standing shock
is capable of reproducing the observed nature of the outburst in 3C~454.3, we have calculated
SEDs and light curves with the Turbulent Extreme Multi-Zone (TEMZ) numerical code developed by
\cite{mar12}. The geometry is sketched in Figure \ref{fig12}. A pressure mismatch at the boundary (with the
interior of the jet being under-pressured) maintains a standing conical shock. A Mach disk, a strong
shock oriented transverse to the jet axis with a radius much less than the cross-sectional jet radius,
truncates the cone. Such a structure can occur when there is a high degree of azimuthal symmetry in the
flow. The Mach disk slows the flow from a highly relativistic velocity to a speed no greater than $c/3$,
where $c$ is the speed of light. This deceleration greatly compresses the magnetic field and density
behind the shock front, so that the Mach disk radiates very efficiently. Since the flow crossing the
conical shock is only modestly decelerated, it remains highly relativistic so that the plasma
receives highly blueshifted emission from the Mach disk. The Mach disk therefore serves as a
very intense source of seed photons that varies as the flow's magnetic field and density of electrons
change.

The TEMZ code divides the emission region beyond the shocks into many cylindrical cells, as illustrated
in Figure \ref{fig12}. Each cell is assigned a uniform magnetic field whose direction is determined randomly.
Relativistic electrons are injected into the cells at the shock front and subsequently lose energy
from synchrotron and inverse Compton radiation as the plasma flows downstream. The jet opening angle
is so small \citep[$\lesssim 1^\circ$;][]{J05} that expansion cooling is negligible across the emission
region. Although seed photons from the hot dust torus are included in the calculation, this source
is negligible relative to the Mach disk synchrotron and first-order synchrotron self-Compton (SSC) radiation
beyond several parsecs from the central engine.
(Second-order SSC within the Mach disk is beyond the Klein-Nishina limit and therefore suppressed.)
The Mach disk spectrum follows that of relativistic electrons in the ``fast cooling'' limit
\citep{se01}, but blueshifted by a factor that depends on the location of each cell.
The energy density of the flow, constant throughout the jet cross-section at a given distance $z$ from
the base, is allowed to vary randomly with time, and therefore with $z$, in a manner that reproduces the
slope of the observed PSD of the $\gamma$-ray emission, $b \sim 1.7$ \citep{abdo2010}. The energy density
is split between the relativistic electrons and magnetic field, with the ratio of the two left as a free
parameter. The calculations include synchrotron self-absorption within the cells and along
the various photon paths.

\begin{figure}
\plottwo{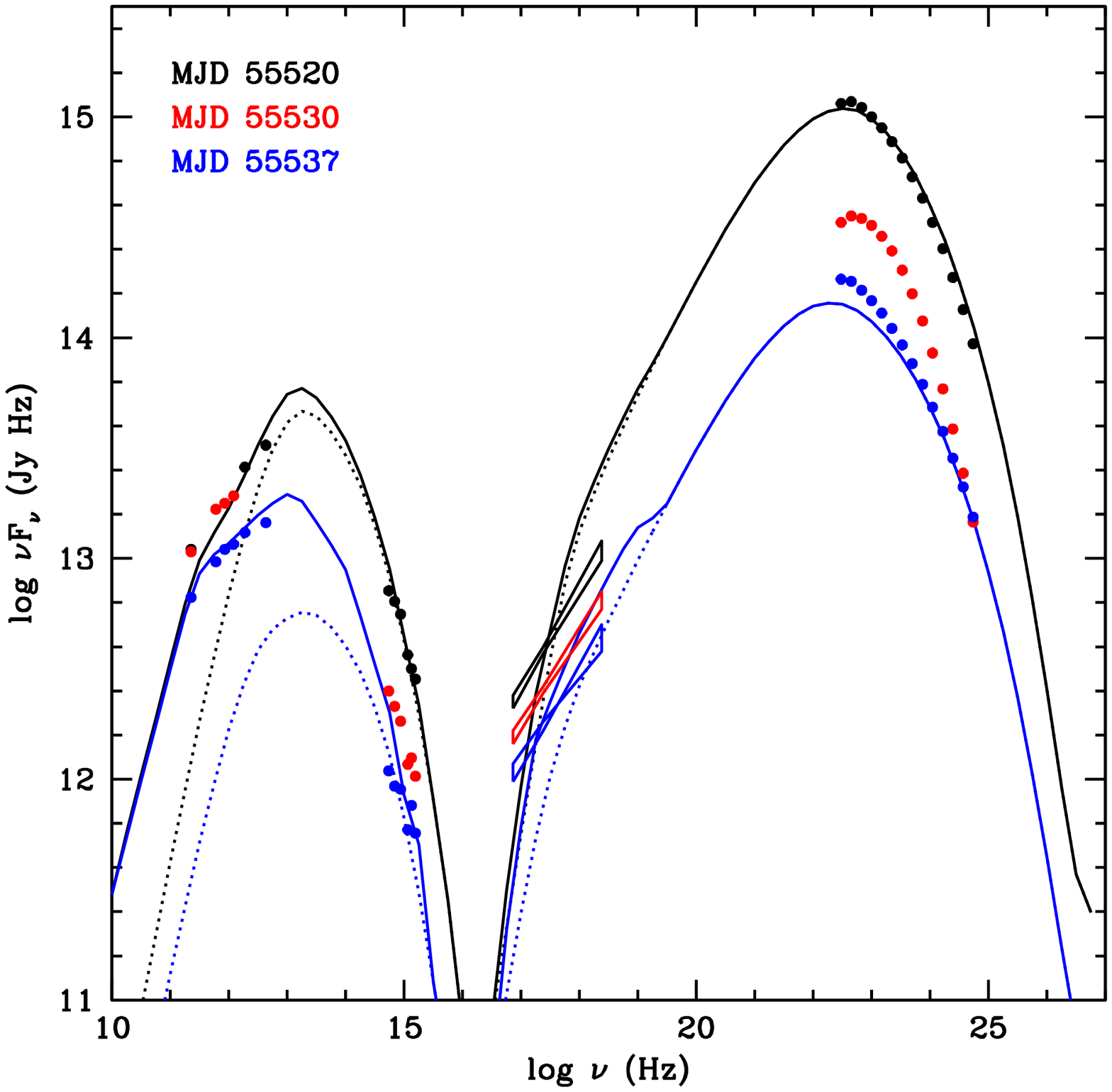}{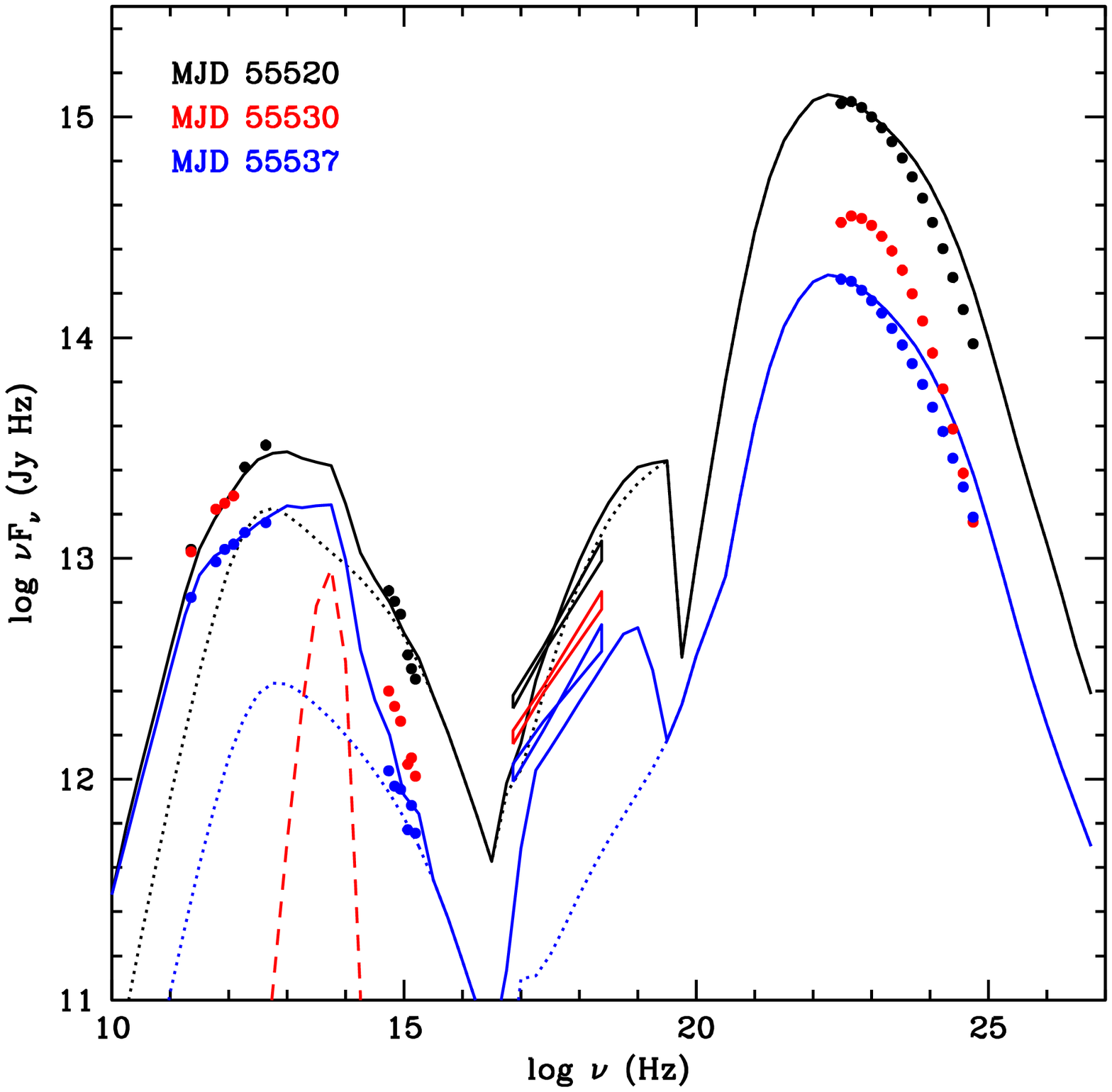}
\caption{Spectral energy distributions from millimeter through $\gamma-$ray  bands at three
epochs. Dotted curves represent simulated SEDs generated by the
TEMZ code at the peak of the light curve (black) and a later time (blue). The dashed red curve represents hot dust. Solid curves add to the
TEMZ curves the more slowly varying emission observed prior to the outburst. {\it Left model:}
seed photons are nonthermal radiation from Mach disk; {\it Right model:} seed photons are
IR radiation from hot dust ($\gamma$-ray) and synchrotron radiation from plasma in cells (X-ray).
\label{fig13}}
\end{figure}

Since there is a strong random component in the TEMZ calculations, we cannot realistically reproduce
the light curve of 3C~454.3 from the results of a reasonable number of accumulated time steps.
[\citet{mar12} presents a sample light curve produced by the code.] Rather,
we have searched enough parameter sets to determine that the simulated SED resembles
that observed at the peak of the outburst, which was artificially induced by 
increasing the 
density of relativistic electrons by a factor of 10 over five time steps. We define this success as the ability to reproduce the
spectral slopes within $\pm 0.1$ and the fluxes within a factor of two at all observed wavelengths. The left panel
of Figure \ref{fig13} presents one of the fits. The numerically induced flare is spread over a number of time
steps owing to the propagation time down the conical shock and light-travel delays from the Mach disk
to the various cells. The version of the code employed here incorporates 270 cells across the jet
plus the Mach disk. Additional cells follow the emission in the flow beyond the conical shock until a rarefaction
is reached, downstream of which expansion cooling is assumed to weaken the emission to a negligible level.

In the model, the steepening of the SED from IR to optical frequencies and toward higher
$\gamma$-ray energies
is a consequence of the electrons radiating in the optical and GeV bands having energies near the
upper end of the electron distribution. These energies can only be maintained close to the injection
site at the shock front, hence the volume filling factor is small relative to the emission by
lower energy electrons at longer wavelengths. The dependence of the filling factor on frequency
is such that the SED can be fit fairly well as a power law over a decade or more in frequency.
Our fits to the SED of 3C~454.3 shown in Figure 13 are only
representative, and surely
not unique, given that we have explored only a relatively narrow range of
free parameters. Fits
to the SEDs on MJD 55520 and 55537 are quite good; the run of TEMZ that
produced these SEDs
missed the intermediate-state SED of MJD 55530 owing to the finite time
steps inherent
to the code, combined with the rapid drop of flux during the declining
stage of the flare. We set the bulk Lorentz factor of the
unshocked flow at 30 and the angle between the jet axis and the line of sight at $0.3^\circ$,
as derived from the VLBA observations \citep{jorstad2012}. In our frame, a given cell of plasma
propagates down the jet at a rate of 0.8 pc day$^{-1}$ until it crosses the shock and decelerates.
This causes very rapid time-scales of variability as turbulent cells enter the shock.
Besides the injected electron energy,
the free parameters (and their adopted values) include the angle of the conical shock to the jet axis
($4^\circ$, which yields a bulk Lorentz factor of the shocked plasma of 13.9 and a mean Doppler
factor of 24), ratio of electron
to magnetic energy density (12 in quiescence), time-averaged pre-shocked
magnetic field strength (0.016 G), cell cross-sectional radius (0.008 pc), and cross-sectional
radius of the Mach disk (0.016 pc). 
The hardening of the $\gamma$-ray spectrum at energies $> 10$ GeV after
the maximum flux of the flare found by Abdo et al. (2011) occurs also in
the SED produced by the TEMZ code: Figure 13 shows a flatter $\gamma$-ray
spectral slope above the peak in the SED during the decaying portion of
the flare. This is expected because lower flux levels reduce the radiative
energy loss rate of the highest-energy electrons. As the flux declines,
more of these electrons can therefore maintain their energies long enough
to make $> 10$ GeV photons via inverse Compton scattering. 

Although the fit to the
observed SED is not unique, other parameter sets providing similar fits do not vary greatly from
these values. The only significant discrepancy with the data
is in the X-ray spectral index $\alpha_{\rm x}$. This is 0.5 in the model, while the observations
give a value of $\sim 0.7$. The spectral index from the model calculation reflects the slope of the electron
energy distribution, which is $-2.0$ at energies below the minimum injected energy owing to
the dependence of the radiative energy loss rate on the square of the energy. The fit to the slopes in
the other observed frequency ranges is quite good, on the other hand.

\subsection{Model Involving Turbulent Plasma Flowing across a Stationary Conical Shock Irradiated
by IR Emission from Hot Dust}

The failure of the TEMZ model with a Mach disk to fit the X-ray data might be the result
of limiting the geometry to uniformity across each cross-section of the pre-shocked flow or some
other simplification. It could also indicate that another source of seed photons dominates the
inverse Compton scattering by electrons in the jet of 3C~454.3. In order to investigate
this possibility, we have also performed a calculation with a different source of seed photons for
inverse Compton scattering: thermal IR emission by dust. We adopt the standard geometry of a torus,
allowing for a patchy structure. Since dust emission in a blazar has been observed directly with a
well-characterized SED only in 4C~21.35 \citep{malmrose11}, we use the IR SED of this blazar as the
standard. The emission is dominated by a blackbody of temperature 1200 K with a luminosity of
$8\times 10^{45}$ erg s$^{-1}$, which is 22\% of the accretion disk luminosity. We note that
the disk luminosity of 3C~454.3 is about 50\% that of 4C~21.35 \citep{vercellone2011}. If
the dust of the latter blazar is confined to a uniform torus, the inner radius is 1.1 pc and
the outer radius 1.9 pc \citep{malmrose11}. The torus might actually be larger if it is
patchy, but at the expense of a lower filling factor of its emission as viewed in the
frame of the jet plasma.

We have included thermal emission from such a dust torus in the TEMZ model after removing the
Mach disk, which is appropriate if substantial azimuthal asymmetries exist in the pressure of
the jet or surrounding confining medium. We used the dust luminosity that is the maximum allowed by the data. The right panel of Figure \ref{fig13} displays SEDs at
two times produced in a simulation that provides a rather good fit to the synchrotron and
$\gamma$-ray SEDs. Inverse Compton scattering of the dust-emitted IR photons greatly under-produces
the X-ray emission. SSC emission (calculated only crudely in the current version of the TEMZ code,
which includes only synchrotron seed photons from a cell and its nearest neighbors) produces
an X-ray flux 
similar to the observed value at the peak of the main flare. 
The parameters used in the model
that differ from those in the Mach disk scenario are: pre-shocked mean magnetic field of 0.01 G,
ratio of electron to magnetic energy density of 100, cell cross-sectional radius of 0.005 pc,
minimum and maximum energy of electrons of $10^3$ and $2\times 10^4$ times $mc^2$, and shock angle of
$3.0^\circ$ to the jet axis. In order to create a sufficient density of seed photons, the dust torus
is allowed to be rather large, with an outer radius of 7 pc and inner radius of 3 pc. We set the areal
covering factor at 0.1 such that the IR luminosity exceeds that expected based on the disk luminosity
by a factor of 3, which allows us to place the upstream end of the conical shock at 1 pc from
the central engine, while the downstream end is at 2.8 pc. If we were to place the emission farther
downstream, the hot dust would need to be spread over an even larger volume to produce enough seed
photons to explain the apparent luminosity of the outburst.

Because the dust torus provides a steady source of seed photons that declines with distance down
the jet, the outward motion of the disturbance can cause the declining portion of a flare. However,
any more rapid fluctuations need to arise from either a change in the electron density or energy
distribution --- in which case the same fluctuation should appear in both the optical and
$\gamma$-ray light curves, since the emission at both regions arises from the same electrons ---
or variations in the Doppler factor from cell to cell. The TEMZ code does not yet
include the latter.

\subsection{Other Possible Sources of Seed Photons}

It is possible that seed photons for inverse Compton scattering could arise from other sources in the nucleus
of the blazar. One of these is nonthermal radiation from a somewhat slower sheath of the jet
that may surround the extremely relativistic ``spine'' \citep[see][and references therein]{mar10}.
This emission might be more local to the primary $\gamma$-ray emission region, perhaps allowing variability on the
observed intra-day time-scales \citep{abdo2011,vercellone2011}. Although the TEMZ code does not yet include
seed photons from the sheath, the resulting emission pattern should be fairly similar to that of the model
that includes seed photons from a Mach disk. The main difference is that emission from the sheath would not
be as concentrated, which would broaden the flare profiles. In addition, the disturbance causing the flare
would propagate at a slower speed in the sheath, so that the amplification of the flare via an increase in
both the number of radiating electrons and density of seed photons in the Mach disk model would not occur.

\citet{lt11} have proposed that emission-line clouds situated near the jet on parsec scales could provide local
sources of seed photons. In this case, the emission-line luminosity would be expected to vary along with the
nonthermal UV continuum, contrary to observations of both 3C~454.3 during the outburst studied here
\citep{jorstad2012} and the $\gamma$-ray bright quasar 4C~21.35 during a multi-waveband outburst in 2010 \citep{smith11}.

\section{Conclusions} 

The variations of flux at millimeter, far-IR, and $\gamma$-ray bands during the outburst of
3C~454.3 from 2010 November to 2011 January were strongly correlated, with the $\gamma$-ray variations
lagging those at 1.3 mm and 160 $\mu$m by $1.0\pm 0.5$ days. Despite the strong correlations, the substructure
of the outburst reveals significant differences at various wavebands. This suggests that physical
substructure is present in the jet, which we interpret as evidence for turbulent processes.

The nearly simultaneous peak of the main flare at 1.3 mm and at $\gamma$-ray energies implies that the 
flaring component of the multiwavelength emission was optically thin and dominated the emission at 
frequencies higher than 43 GHz. At 1.3 mm, the core region and knot K09 (ejected
in late 2009) were the main emitters. Analysis of the light curve of the two components leads to the conclusion
that the core region was responsible for the 1.3 mm outburst, while the flux of K09 declined during the event.
The lack of a significant time delay between the flare at 1.3 mm
and $\gamma$-ray energies requires that the flare takes place in a region that is transparent
at 1.3 mm. In support of this, the core region brightened by 70\% at 7 mm during
the early stages of the outburst bracketing the main flare. The data therefore favor a location of the
outburst within the parsec-scale core rather than within the BLR. This conflicts with the model proposed
by \citet{vercellone2011}, who place the flare in a
region $3.6\times 10^{16}$ cm in radius only 0.05 pc from the central engine.
In addition, later VLBA observations have identified a bright superluminal knot
that was blended with the core during the outburst. The disturbance creating the flare therefore appears
to have created the new knot. 

Our {\it Herschel} far-IR observations clearly define the low-frequency side of the maximum in the 
synchrotron SED, from which we infer that the SED peaked at a wavelength shorter than 70 $\mu$m. Two weeks before the flare, IRTF observations located the peak at 11 $\mu$m. The SED of the main flare
did not shift toward lower frequencies as rapidly as expected in shock or expanding plasmoid models.
This leads us to interpret the outburst in terms of a standing shock in the jet, across which turbulent
plasma flows. A major increase in the energy density of the plasma causes an outburst as it crosses
the shock, which has a conical structure. If the conical shock terminates in a transversely oriented
Mach disk, a strong shock that decelerates and compresses the flow greatly, emission from the Mach disk can
provide the main source of seed photons for inverse Compton scattering to generate the X-ray and
$\gamma$-ray emission. The TEMZ code  produces model SEDs, following this scenario, that match the millimeter to optical and $\gamma$-ray spectra quite well although the observed X-ray spectrum is somewhat steeper than in the model calculations.
 Alternatively, if asymmetries in the jet flow prevent the formation of
a Mach disk, thermal IR emission from hot dust can provide the requisite seed photons. 
However, the dust would need to have a luminosity $\sim 1\times 10^{46}$ erg s$^{-1}$ --- half that
of the accretion disk --- and would need to be very patchy. 
In the TEMZ model, the volume filling factor of the emission is inversely
related to the frequency of observation. Because of this, the average
degree of linear polarization, as well as the level of variability of both
the flux and polarization, should increase with frequency in a manner that
is quantitatively related to the amount of steepening of the spectral
slope toward high frequencies \citep{mj10}. Another
prediction of the model is that details of the flare profiles should
change randomly from one flare to another.

Our observations of 3C~454.3 therefore present significant challenges to all theoretical models
that have been proposed thus far. Since the polarization properties of blazars require irregularities
in the magnetic field that are most easily understood as the effects of turbulence, scenarios
similar to the TEMZ model appear to be necessary to reproduce the observed multi-waveband behavior
of blazars. Whether this class of models can succeed in doing so requires further development of such models
as well as more comprehensive observations of distinct outbursts in blazars.

\acknowledgments

We are grateful to the {\it Herschel} Director Goran Pilbratt, {\it Swift} Principal Investigator Neil Gehrels,  Herschel ESA staff astronomers Mark Kidger, Pedro Garcia-Lario, and Herschel NASA staff astronomers Babar Ali, Nicolas Billot, Phil Appleton and Bernhard Schulz. We also thank the Herschel spacecraft schedulers in Munich for setting up our monitoring program. 

A. Wehrle acknowledges Guest Investigator support from NASA via Herschel RSA 1427799 and Fermi Guest
Investigator grant NNX11AAO85G. The BU group acknowledges support by NASA under Fermi Guest
Investigator grants  NNX08AV65G, NNX10AO59G, NNX10AU15G, and NNX11AQ03G and Swift Guest Investigator grant
NNX10AF88G.
The Submillimeter Array is a joint project between the Smithsonian Astrophysical Observatory and the Academia Sinica Institute of Astronomy and Astrophysics and is funded by the Smithsonian Institution and the Academia Sinica.
The VLBA is an instrument of the National Radio Astronomy Observatory. The National Radio Astronomy Observatory is a facility of the National Science
Foundation operated under cooperative agreement by Associated Universities, Inc. This research has made use of the
 XRT Data Analysis Software (XRTDAS) developed under the responsibility
 of the ASI Science Data Center (ASDC), Italy.
Swift at PSU is supported by NASA contract NAS5-00136.
I. Agudo acknowledges funding support from the regional government of Andaluc\'{i}a and the ``Ministerio de Ciencia e Innovaci\'{o}n'' of Spain through grants P09-FQM-4784 and AYA2010-14844, respectively.

{\it Facilities:} \facility{Herschel}, \facility{Fermi}, \facility{Swift}, \facility{SMA}, \facility{IRTF}, \facility{VLBA}.

\clearpage

\clearpage

\begin{deluxetable}{ccccccccc}
\tablecaption{SMA and Herschel Data}
\tablewidth{0pt}
\singlespace
\tablecolumns{9}
\tabletypesize{\small}
\tablehead{
             \colhead{Date}
             & \colhead{MJD}
             & \colhead{1.3 mm}
             & \colhead{1.3 mm}
             & \colhead{500 \micron}
             & \colhead{350 \micron}
             & \colhead{250 \micron}
             & \colhead{160 \micron}
             & \colhead{70 \micron}
             \\
             \colhead{ }
             & \colhead{ }
             & \colhead{Flux}
             & \colhead{Flux}
             & \colhead{Flux}
             & \colhead{Flux}
             & \colhead{Flux}
             & \colhead{Flux}
             & \colhead{Flux}
             \\
             \colhead{ }
             & \colhead{ }
             & \colhead{Density}
             & \colhead{Density Error}
             & \colhead{Density\tablenotemark{a}}
             & \colhead{Density\tablenotemark{a}}
             & \colhead{Density\tablenotemark{a}}
             & \colhead{Density\tablenotemark{a}}
             & \colhead{Density\tablenotemark{a}}
             \\

	        \colhead{    } 
	       & \colhead{    } 
	      & \colhead{[Jy]} 
             & \colhead{[Jy]}
	      & \colhead{[Jy]} 
             & \colhead{[Jy]}
	       & \colhead{[Jy]    } 
	      & \colhead{[Jy]} 
             & \colhead{[Jy]}\\
}
\startdata
19-Nov-2010	 & 	55519	 & 	48.8	 & 	2.4	 & 	 \nodata 	 & 	 \nodata	 & 	 \nodata	 & 	13.8	 & 	7.61	\\
20-Nov-2010	 & 	55520	 & 	51.8	 & 	4.3	 & 	 \nodata	 & 	 \nodata	 & 	 \nodata	 & 	 \nodata	 & 	 \nodata	\\
21-Nov-2010	 & 	55521	 & 	45.5	 & 	2.3	 & 	 \nodata	 & 	 \nodata	 & 	 \nodata	 & 	 \nodata	 & 	 \nodata	\\
22-Nov-2010	 & 	55522	 & 	42.9	 & 	2.1	 & 	 \nodata	 & 	 \nodata	 & 	 \nodata	 & 	 \nodata	 & 	 \nodata	\\
23-Nov-2010	 & 	55523	 & 	41.7	 & 	2.1	 & 	23.48	 & 	18.63	 & 	14.81	 & 	 \nodata	 & 	 \nodata	\\
24-Nov-2010	 & 	55524	 & 	41.8	 & 	3.1	 & 	 \nodata	 & 	 \nodata	 & 	 \nodata	 & 	 \nodata	 & 	 \nodata	\\
25-Nov-2010	 & 	55525	 & 	 \nodata	 & 	 \nodata	 & 	 \nodata	 & 	 \nodata	 & 	 \nodata	 & 	 \nodata	 & 	 \nodata	\\
26-Nov-2010	 & 	55526	 & 	 \nodata	 & 	 \nodata	 & 	 \nodata	 & 	 \nodata	 & 	 \nodata	 & 	 \nodata	 & 	 \nodata	\\
\enddata
\tablenotetext{a}{The errors are 5\% at 500, 350 and 250 \micron, 4.15\% at 160 \micron ~and 2.64\% at 70 \micron .}
\end{deluxetable}
\clearpage

\begin{deluxetable}{cccccc}
\tablecaption{Swift UVOT V and B Band Photometry}
\tablewidth{0pt}
\singlespace
\tablecolumns{6}
\tabletypesize{\small}
\tablehead{
              \colhead{MJD}
             & \colhead{V band}
             & \colhead{V band}
             & \colhead{MJD}
             & \colhead{B band}
             & \colhead{B band}
             \\
             \colhead{ }
             & \colhead{Flux}
             & \colhead{Flux}
             & \colhead{ }
             & \colhead{Flux}
             & \colhead{Flux}
             \\
             \colhead{ }
                  & \colhead{Density}
             & \colhead{Density Error}
        & \colhead{ }
            & \colhead{Density}
             & \colhead{Density Error}
             \\
	        \colhead{    } 
	      & \colhead{[mJy]} 
	       & \colhead{[mJy]    } 
	       & \colhead{    } 
	      & \colhead{[mJy]} 
             & \colhead{[mJy]}\\
}
\startdata
55471.4764	 & 	3.67	 & 	0.13	 & 	55471.4739	 & 	3.01	 & 	0.07	 \\
55477.8188	 & 	3.13	 & 	0.08	 & 	55477.8133	 & 	2.34	 & 	0.04	 \\
55491.7490	 & 	2.61	 & 	0.07	 & 	55487.5339	 & 	2.56	 & 	0.04	 \\
55498.5102	 & 	4.00	 & 	0.08	 & 	55491.7421	 & 	2.13	 & 	0.04	 \\
55502.4454	 & 	10.49	 & 	0.12	 & 	55498.5036	 & 	2.98	 & 	0.04	 \\
55504.1962	 & 	6.59	 & 	0.12	 & 	55502.4385	 & 	7.30	 & 	0.06	 \\
55504.6523	 & 	7.14	 & 	0.13	 & 	55504.1914	 & 	4.97	 & 	0.06	 \\
55505.1875	 & 	7.22	 & 	0.12	 & 	55504.6474	 & 	5.17	 & 	0.06	 \\
\enddata
\end{deluxetable}
\clearpage


\begin{deluxetable}{cccccc}
\tablecaption{Swift UVOT U and UVW1 Band Photometry}
\tablewidth{0pt}
\singlespace
\tablecolumns{6}
\tabletypesize{\small}
\tablehead{
              \colhead{MJD}
             & \colhead{U band}
             & \colhead{U band}
             & \colhead{MJD}
             & \colhead{ UVW1 band}
             & \colhead{UVW1 band}
             \\
             \colhead{ }
             & \colhead{Flux}
             & \colhead{Flux}
             & \colhead{ }
             & \colhead{Flux}
             & \colhead{Flux}
             \\
             \colhead{ }
             & \colhead{Density}
             & \colhead{Density Error}
             & \colhead{ }
            & \colhead{Density}
             & \colhead{Density Error}
             \\
	        \colhead{    } 
	      & \colhead{[mJy]} 
	       & \colhead{[mJy]    } 
	       & \colhead{    } 
	      & \colhead{[mJy]} 
             & \colhead{[mJy]}\\
}
\startdata
55471.4734	 & 	2.45	 & 	0.05	 & 	55471.4726	 & 	1.63	 & 	0.03	 \\ 
55477.8122	 & 	1.94	 & 	0.03	 & 	55477.8106	 & 	1.33	 & 	0.02	 \\ 
55487.5328	 & 	1.82	 & 	0.03	 & 	55487.5312	 & 	1.31	 & 	0.02	 \\ 
55491.7407	 & 	1.79	 & 	0.03	 & 	55491.7386	 & 	1.31	 & 	0.02	 \\ 
55498.5022	 & 	2.29	 & 	0.03	 & 	55498.5002	 & 	1.49	 & 	0.02	 \\ 
55502.4371	 & 	5.07	 & 	0.04	 & 	55502.4350	 & 	2.75	 & 	0.02	 \\ 
55504.1904	 & 	3.43	 & 	0.04	 & 	55503.1719	 & 	2.40	 & 	0.02	 \\ 
55504.6464	 & 	3.58	 & 	0.04	 & 	55503.7221	 & 	2.09	 & 	0.01	 \\
\enddata
\end{deluxetable}
\clearpage

\begin{deluxetable}{cccccc}
\tablecaption{Swift UVOT UVM2 and UVW2 Band Photometry}
\tablewidth{0pt}
\singlespace
\tablecolumns{6}
\tabletypesize{\small}
\tablehead{
              \colhead{MJD}
             & \colhead{UVM2 band}
             & \colhead{UVM2 band}
             & \colhead{MJD}
             & \colhead{UVW2 band}
             & \colhead{UVW2 band}
             \\
             \colhead{ }
             & \colhead{Flux}
             & \colhead{Flux}
             & \colhead{ }
             & \colhead{Flux}
             & \colhead{Flux}
             \\
             \colhead{ }
                  & \colhead{Density}
             & \colhead{Density Error}
        & \colhead{ }
            & \colhead{Density}
             & \colhead{Density Error}
             \\
	        \colhead{    } 
	      & \colhead{[mJy]} 
	       & \colhead{[mJy]    } 
	       & \colhead{    } 
	      & \colhead{[mJy]} 
             & \colhead{[mJy]}\\
}
\startdata
55471.4773	 & 	1.45	 & 	0.03	 &   	55471.4752	 & 	1.22	 & 	0.02	 \\ 
55477.8208	 & 	1.30	 & 	0.02	 &   	55477.8160	 & 	1.02	 & 	0.01	 \\ 
55491.7516	 & 	1.21	 & 	0.02	 &   	55487.5364	 & 	0.96	 & 	0.01	 \\ 
55498.5128	 & 	1.30	 & 	0.02	 &   	55491.7455	 & 	0.95	 & 	0.01	 \\ 
55502.4481	 & 	2.28	 & 	0.02	 &   	55498.5069	 & 	0.96	 & 	0.01	 \\ 
55504.1981	 & 	1.70	 & 	0.02	 &   	55502.4419	 & 	1.71	 & 	0.01	 \\ 
55504.6542	 & 	1.90	 & 	0.02	 &   	55504.1938	 & 	1.36	 & 	0.01	 \\ 
55505.1895	 & 	1.87	 & 	0.02	 &   	55504.6499	 & 	1.44	 & 	0.01	 \\
\enddata
\end{deluxetable}
\clearpage

\begin{deluxetable}{cccccc}
\tablecaption{Swift X-ray Photometry and Photon Indices}
\tablewidth{0pt}
\singlespace
\tablecolumns{6}
\tabletypesize{\small}
\tablehead{
              \colhead{MJD}
             & \colhead{X-ray}
             & \colhead{X-ray}
             & \colhead{X-ray}
             & \colhead{Photon}
             & \colhead{Photon}
             \\
             \colhead{ }
             & \colhead{Flux}
             & \colhead{Flux Lower Error}
             & \colhead{ Flux Upper Error}
             & \colhead{Index}
             & \colhead{Index Error}
             \\
	        \colhead{    } 
	      & \colhead{[ergs cm$^{-2}$ s$^{-1}$]} 
	       & \colhead{[ergs cm$^{-2}$ s$^{-1}$] } 
	       & \colhead{[ergs cm$^{-2}$ s$^{-1}$]} 
	      & \colhead{} 
             & \colhead{}\\
}
\startdata

55457.4011	 & 	4.21E-11	 & 	-3.1E-12	 & 	3.2E-12	 & 	1.64	 & 	0.07	 \\ 
55463.6294	 & 	4.26E-11	 & 	-2.8E-12	 & 	2.9E-12	 & 	1.78	 & 	0.08	 \\ 
55471.4720	 & 	3.15E-11	 & 	-2.3E-12	 & 	2.8E-12	 & 	1.65	 & 	0.09	 \\ 
55477.8094	 & 	3.99E-11	 & 	-2.7E-12	 & 	3.1E-12	 & 	1.67	 & 	0.07	 \\ 
55487.5300	 & 	5.89E-11	 & 	-4.6E-12	 & 	4.8E-12	 & 	1.67	 & 	0.08	 \\ 
55491.7372	 & 	6.53E-11	 & 	-3.1E-12	 & 	3.3E-12	 & 	1.64	 & 	0.05	 \\ 
55498.4989	 & 	5.91E-11	 & 	-3.0E-12	 & 	3.2E-12	 & 	1.67	 & 	0.06	 \\ 
55502.4336	 & 	5.39E-11	 & 	-1.9E-12	 & 	1.7E-12	 & 	1.65	 & 	0.04	 \\
\enddata
\end{deluxetable}
\clearpage
\begin{deluxetable}{cccccccc}
\tablecaption{Fermi LAT Data}
\tablewidth{0pt}
\singlespace
\tablecolumns{8}
\tabletypesize{\small}
\tablehead{
              \colhead{MJD}
             & \colhead{Photon}
             & \colhead{Photon}
             & \colhead{$\alpha$\tablenotemark{a}}
             & \colhead{$\beta$\tablenotemark{a}}
             & \colhead{Break\tablenotemark{a}}
             & \colhead{Test}
             \\
             \colhead{ }
             & \colhead{Flux}
             & \colhead{Flux Error}
             & \colhead{ }
             & \colhead{}
             & \colhead{Energy}
             & \colhead{Statistic}
             \\
	        \colhead{    } 
	      & \colhead{[ph cm$^{-2}$ s$^{-1}$]} 
	       & \colhead{[ph cm$^{-2}$ s$^{-1}$] } 
	       & \colhead{} 
	      & \colhead{} 
             & \colhead{[GeV]}
             & \colhead{}\\
}
\startdata
55480.5	 & 	3.02E-06	 & 	4.9E-07	 & 	2.21	 & 	0.338	 & 	296	 & 	271	 & 		 \\ 
55481.5	 & 	2.91E-06	 & 	4.5E-07	 & 	2.06	 & 	0.187	 & 	293	 & 	256	 & 		 \\ 
55482.5	 & 	3.41E-06	 & 	4.9E-07	 & 	2.44	 & 	0.317	 & 	298	 & 	259	 & 		 \\ 
55483.5	 & 	3.44E-06	 & 	5.0E-07	 & 	2.45	 & 	0.117	 & 	298	 & 	217	 & 		 \\ 
55484.5	 & 	3.55E-06	 & 	4.6E-07	 & 	2.34	 & 	0.248	 & 	302	 & 	400	 & 		 \\ 
55485.5	 & 	3.06E-06	 & 	5.2E-07	 & 	2.14	 & 	0.107	 & 	290	 & 	342	 & 		 \\ 
55486.5	 & 	3.43E-06	 & 	4.5E-07	 & 	2.40	 & 	0.100	 & 	288	 & 	321	 & 		 \\ 
55487.5	 & 	3.28E-06	 & 	5.4E-07	 & 	2.39	 & 	0.155	 & 	306	 & 	269	 & 		 \\
\enddata
\tablenotetext{a}{The errors in $\alpha$ are estimated at 6\% and for $\beta$, 15\% when the photon flux exceeds $10^{-5}$~ $ph/s/cm^{2}$ and 35\% below that level, and for the break energy at 30\%.  See text for explanation of correlated errors.}
\end{deluxetable}
\clearpage

\begin{deluxetable}{cccccccccccccc}
\tablecaption{SED Data in Millimeter to Ultraviolet Bands}
\rotate
\tablewidth{0pt}
\singlespace
\tablecolumns{13}
\tabletypesize{\scriptsize}
\tablehead{
              \colhead{MJD}
             & \colhead{2.25 E11}
             & \colhead{6.00 E11}
             & \colhead{8.57 E11}
             & \colhead{1.20 E12}
             & \colhead{1.88 E12}
             & \colhead{4.29 E12}
             & \colhead{5.49 E14}
             & \colhead{6.83 E14}
              & \colhead{8.66 E14}
             & \colhead{1.15 E15}
             & \colhead{1.34 E15}
             & \colhead{1.56 E15}            
             \\
             \colhead{ }
             & \colhead{(225 GHz)}
             & \colhead{500 \micron}
             & \colhead{350 \micron }
             & \colhead{250 \micron}
             & \colhead{160 \micron}
             & \colhead{70 \micron}
             & \colhead{V}
              & \colhead{B}
               & \colhead{U} 
               & \colhead{UVW1} 
                           & \colhead{UVM2}
                           & \colhead{UVW2}  
                                     
             \\
	        \colhead{    } 
	      & \colhead{[Jy Hz]} 
	      & \colhead{[Jy Hz]} 
	      & \colhead{[Jy Hz]} 
	      & \colhead{[Jy Hz]} 
	      & \colhead{[Jy Hz]} 
	      & \colhead{[Jy Hz]} 
	      & \colhead{[Jy Hz]} 
	      & \colhead{[Jy Hz]} 
	      & \colhead{[Jy Hz]} 
	      & \colhead{[Jy Hz]} 
	      & \colhead{[Jy Hz]} 
	      & \colhead{[Jy Hz]} 
              & \colhead{}\\
}
\startdata

55519	 & 	1.10E+13	 & 	 \nodata	 & 	 \nodata	 & 	 \nodata	 &	2.59E+13	 &	3.26E+13	 &	8.34E+12	 &	7.60E+12	 &	6.78E+12	 &	4.85E+12	 &	4.37E+12	 &	4.04E+12	 \\ 
55520	 & 	1.16E+13	 & 	 \nodata	 & 	 \nodata	 & 	 \nodata	 &	 \nodata	 &	 \nodata	 &	 \nodata	 &	 \nodata	 &	8.49E+12	 &	 \nodata	 &	 \nodata	 &	 \nodata	 \\ 
55521	 & 	1.02E+13	 & 	 \nodata	 & 	 \nodata	 & 	 \nodata	 &	 \nodata	 &	 \nodata	 &	 \nodata	 &	 \nodata	 &	 \nodata	 &	 \nodata	 &	 \nodata	 &	3.18E+12	 \\ 
55522	 & 	9.66E+12	 & 	 \nodata	 & 	 \nodata	 & 	 \nodata	 &	 \nodata	 &	 \nodata	 &	 \nodata	 &	 \nodata	 &	 \nodata	 &	 \nodata	 &	3.22E+12	 &	 \nodata	 \\ 
55523	 & 	9.39E+12	 & 	1.41E+13	 & 	1.60E+13	 & 	1.78E+13	 &	 \nodata	 &	 \nodata	 &	 \nodata	 &	 \nodata	 &	 \nodata	 &	3.63E+12	 &	 \nodata	 &	 \nodata	 \\ 
55524	 & 	9.40E+12	 & 	 \nodata	 & 	 \nodata	 & 	 \nodata	 &	 \nodata	 &	 \nodata	 &	 \nodata	 &	 \nodata	 &	3.55E+12	 &	 \nodata	 &	 \nodata	 &	 \nodata	 \\ 
55525	 & 	 \nodata	 & 	 \nodata	 & 	 \nodata	 & 	 \nodata	 &	 \nodata	 &	 \nodata	 &	3.85E+12	 &	3.63E+12	 &	3.21E+12	 &	2.59E+12	 &	2.44E+12	 &	2.30E+12	 \\ 
55526	 & 	 \nodata	 & 	 \nodata	 & 	 \nodata	 & 	 \nodata	 &	 \nodata	 &	 \nodata	 &	3.75E+12	 &	3.59E+12	 &	3.13E+12	 &	2.37E+12	 &	2.48E+12	 &	2.24E+12	 \\
\enddata
\end{deluxetable}
\clearpage

\begin{deluxetable}{cccccc}
\tablecaption{SED Data in X-ray and $\gamma$-ray Bands}
\tablewidth{0pt}
\singlespace
\tablecolumns{6}
\tabletypesize{\small}
\tablehead{
              \colhead{MJD}
             & \colhead{7.25 E16 Hz}
             & \colhead{9.67 E17 Hz}
             & \colhead{2.42E18 Hz}
             & \colhead{2.43 E22 Hz}
             & \colhead{7.25 E22 Hz}     
             \\
             \colhead{ }
             & \colhead{(0.3 keV)}
             & \colhead{(4 keV)}
             & \colhead{(10 keV) }
             & \colhead{(100 MeV)}
             & \colhead{(300 MeV)}
             \\
	        \colhead{    } 
	      & \colhead{[Jy Hz]} 
	      & \colhead{[Jy Hz]} 
	      & \colhead{[Jy Hz]} 
	      & \colhead{[Jy Hz]} 
	      & \colhead{[Jy Hz]} 
\\
}
\startdata

55519	 &	1.05E+12	 &	6.28E+12	 &	9.22E+12	 &	1.22E+15	 &	1.08E+15	 \\ 
55520	 &	1.05E+12	 &	6.51E+12	 &	9.82E+12	 &	9.30E+14	 &	9.81E+14	 \\ 
55521	 &	7.00E+11	 &	3.98E+12	 &	5.66E+12	 &	4.41E+14	 &	4.81E+14	 \\ 
55522	 &	6.15E+11	 &	3.77E+12	 &	5.65E+12	 &	3.54E+14	 &	3.89E+14	 \\ 
55523	 &	6.32E+11	 &	3.60E+12	 &	5.13E+12	 &	4.16E+14	 &	3.64E+14	 \\ 
55524	 &	5.70E+11	 &	3.30E+12	 &	4.76E+12	 &	3.41E+14	 &	3.71E+14	 \\ 
55525	 &	5.41E+11	 &	3.04E+12	 &	4.30E+12	 &	3.44E+14	 &	3.24E+14	 \\ 
55526	 &	6.18E+11	 &	3.56E+12	 &	5.11E+12	 &	4.67E+14	 &	4.78E+14	 \\
\enddata
\end{deluxetable}
\clearpage
\end{document}